\documentclass[twocolumn]{aastex631}

\received{}
\revised{}
\accepted{}

\submitjournal{ApJ}

\shorttitle{ACA CO(2-1) Mapping of M33}
\shortauthors{Muraoka et al.}
\graphicspath{{./}{figures/}}
\begin{document}

\title{ACA CO($J=2-1$) Mapping of the Nearest Spiral Galaxy M33. I. Initial Results and Identification of Molecular Clouds}

\author[0000-0002-3373-6538]{Kazuyuki Muraoka}
\affiliation{Department of Physics, Graduate School of Science, Osaka Metropolitan University, 1-1 Gakuen-cho, Naka-ku, Sakai, Osaka 599-8531, Japan}

\collaboration{20}{(AAS Journals Data Editors)}

\author[0000-0002-4098-8100]{Ayu Konishi}
\affiliation{Department of Physics, Graduate School of Science, Osaka Metropolitan University, 1-1 Gakuen-cho, Naka-ku, Sakai, Osaka 599-8531, Japan}

\author[0000-0002-2062-1600]{Kazuki Tokuda}
\affiliation{Department of Earth and Planetary Sciences, Faculty of Science, Kyushu University, Nishi-ku, Fukuoka 819-0395, Japan}
\affiliation{National Astronomical Observatory of Japan, National Institutes of Natural Science, 2-21-1 Osawa, Mitaka, Tokyo 181-8588, Japan}
\affiliation{Department of Physics, Graduate School of Science, Osaka Metropolitan University, 1-1 Gakuen-cho, Naka-ku, Sakai, Osaka 599-8531, Japan}

\author[0000-0002-3499-9460]{Hiroshi Kondo}
\affiliation{Department of Physics, Graduate School of Science, Osaka Metropolitan University, 1-1 Gakuen-cho, Naka-ku, Sakai, Osaka 599-8531, Japan}

\author[0000-0001-8187-7856]{Rie E. Miura}
\affiliation{Departamento de Fisica Teorica y del Cosmos, Campus de Fuentenueva, Universidad de Granada, E18071-Granada, Spain}
\affiliation{National Astronomical Observatory of Japan, National Institutes of Natural Science, 2-21-1 Osawa, Mitaka, Tokyo 181-8588, Japan}

\author[0000-0001-9016-2641]{Tomoka Tosaki}
\affiliation{Joetsu University of Education, Yamayashiki-machi, Joetsu, Niigata 943-8512, Japan}

\author{Sachiko Onodera}
\affiliation{Meisei University, 2-1-1 Hodokubo, Hino, Tokyo 191-0042, Japan}

\author[0000-0002-1234-8229]{Nario Kuno}
\affiliation{Division of Physics, Faculty of Pure and Applied Sciences, University of Tsukuba, 1-1-1 Tennodai, Tsukuba, Ibaraki 305-8577, Japan}
\affiliation{Tomonaga Center for the History of the Universe, University of Tsukuba, Tsukuba, Ibaraki 305-8571, Japan}

\author[0000-0003-3990-1204]{Masato I. N. Kobayashi}
\affiliation{National Astronomical Observatory of Japan, National Institutes of Natural Science, 2-21-1 Osawa, Mitaka, Tokyo 181-8588, Japan}
\affiliation{I. Physikalisches Institut, Universit\"{a}t zu K\"{o}ln, Z\"{u}lpicher Str 77, D-50937 K\"{o}ln, Germany}

\author[0000-0002-2794-4840]{Kisetsu Tsuge}
\affiliation{Department of Physics, The University of Tokyo, 7-3-1 Hongo, Bunkyo-ku, Tokyo 113-0033, Japan}

\author[0000-0003-2062-5692]{Hidetoshi Sano}
\affiliation{Faculty of Engineering, Gifu University, 1-1 Yanagido, Gifu 501-1193, Japan}

\author{Naoya Kitano}
\affiliation{Department of Physics, Graduate School of Science, Osaka Metropolitan University, 1-1 Gakuen-cho, Naka-ku, Sakai, Osaka 599-8531, Japan}

\author[0000-0002-6375-7065]{Shinji Fujita}
\affiliation{Institute of Astronomy, The University of Tokyo, 2-21-1, Osawa, Mitaka, Tokyo 181-0015, Japan}
\affiliation{Department of Physics, Graduate School of Science, Osaka Metropolitan University, 1-1 Gakuen-cho, Naka-ku, Sakai, Osaka 599-8531, Japan}

\author[0000-0003-0732-2937]{Atsushi Nishimura}
\affiliation{National Astronomical Observatory of Japan, National Institutes of Natural Science, 2-21-1 Osawa, Mitaka, Tokyo 181-8588, Japan}

\author[0000-0001-7826-3837]{Toshikazu Onishi}
\affiliation{Department of Physics, Graduate School of Science, Osaka Metropolitan University, 1-1 Gakuen-cho, Naka-ku, Sakai, Osaka 599-8531, Japan}

\author[0000-0003-1549-6435]{Kazuya Saigo}
\affiliation{Graduate School of Science and Engineering, Kagoshima University, 1-21-40 Korimoto Kagoshima-city Kagoshima, 890-0065, Japan}

\author[0000-0002-1865-4729]{Rin I. Yamada}
\affiliation{Department of Physics, Nagoya University, Chikusa-ku, Nagoya 464-8602, Japan}

\author{Fumika Demachi}
\affiliation{Department of Physics, Nagoya University, Chikusa-ku, Nagoya 464-8602, Japan}

\author[0000-0002-1411-5410]{Kengo Tachihara}
\affiliation{Department of Physics, Nagoya University, Chikusa-ku, Nagoya 464-8602, Japan}

\author[0000-0002-8966-9856]{Yasuo Fukui}
\affiliation{Department of Physics, Nagoya University, Chikusa-ku, Nagoya 464-8602, Japan}
\affiliation{Institute for Advanced Research, Nagoya University, Furo-cho, Chikusa-ku, Nagoya 464-8601, Japan}

\author[0000-0001-7813-0380]{Akiko Kawamura}
\affiliation{National Astronomical Observatory of Japan, National Institutes of Natural Science, 2-21-1 Osawa, Mitaka, Tokyo 181-8588, Japan}

%% Note that the \and command from previous versions of AASTeX is now
%% depreciated in this version as it is no longer necessary. AASTeX 
%% automatically takes care of all commas and "and"s between authors names.

%% AASTeX 6.31 has the new \collaboration and \nocollaboration commands to
%% provide the collaboration status of a group of authors. These commands 
%% can be used either before or after the list of corresponding authors. The
%% argument for \collaboration is the collaboration identifier. Authors are
%% encouraged to surround collaboration identifiers with ()s. The 
%% \nocollaboration command takes no argument and exists to indicate that
%% the nearby authors are not part of surrounding collaborations.

%% Mark off the abstract in the ``abstract'' environment. 
\begin{abstract}
We present the results of ALMA-ACA 7\,m-array observations in $^{12}$CO($J=2-1$), $^{13}$CO($J=2-1$), and C$^{18}$O($J=2-1$) line emission
toward the molecular-gas disk in the Local Group spiral galaxy M33 at an angular resolution of 7\farcs31 $\times$ 6\farcs50 (30\,pc $\times$ 26\,pc).
We combined the ACA 7\,m-array $^{12}$CO($J=2-1$) data with the IRAM 30\,m data to compensate for emission from diffuse molecular-gas components.
The ACA+IRAM combined $^{12}$CO($J=2-1$) map clearly depicts the cloud-scale molecular-gas structure over the M33 disk.
Based on the ACA+IRAM $^{12}$CO($J=2-1$) cube data, we cataloged 848 molecular clouds with a mass range from $10^3$\,$M_{\odot}$ to $10^6$\,$M_{\odot}$.
We found that high-mass clouds ($\geq 10^5\,M_{\odot}$) tend to associate with the $8\,\mu$m-bright sources in the spiral arm region,
while low-mass clouds ($< 10^5\,M_{\odot}$) tend to be apart from such $8\,\mu$m-bright sources and to exist in the inter-arm region.
We compared the cataloged clouds with GMCs observed by the IRAM 30\,m telescope at 49\,pc resolution \citep[IRAM GMC:][]{corbelli2017},
and found that a small IRAM GMC is likely to be identified as a single molecular cloud even in ACA+IRAM CO data, while a large IRAM GMC can be resolved into multiple ACA+IRAM clouds.
The velocity dispersion of a large IRAM GMC is mainly dominated by the line-of-sight velocity difference between small clouds inside the GMC rather than the internal cloud velocity broadening.
\end{abstract}

%% Keywords should appear after the \end{abstract} command. 
%% The AAS Journals now uses Unified Astronomy Thesaurus concepts:
%% https://astrothesaurus.org
%% You will be asked to selected these concepts during the submission process
%% but this old "keyword" functionality is maintained in case authors want
%% to include these concepts in their preprints.

%% From the front matter, we move on to the body of the paper.
%% Sections are demarcated by \section and \subsection, respectively.
%% Observe the use of the LaTeX \label
%% command after the \subsection to give a symbolic KEY to the
%% subsection for cross-referencing in a \ref command.
%% You can use LaTeX's \ref and \label commands to keep track of
%% cross-references to sections, equations, tables, and figures.
%% That way, if you change the order of any elements, LaTeX will
%% automatically renumber them.
%%
%% We recommend that authors also use the natbib \citep
%% and \citet commands to identify citations.  The citations are
%% tied to the reference list via symbolic KEYs. The KEY corresponds
%% to the KEY in the \bibitem in the reference list below. 

\section{Introduction} \label{sec:intro}

The interstellar medium (ISM) is one of the crucial components in galaxies because stars are formed by the contraction of molecular ISM.
In the Milky Way (MW), a large fraction of molecular ISM is in the form of giant molecular clouds \citep[GMCs:][]{sanders1985}, whose typical size and mass are a few $\times$ $10 - 100$\,pc and $10^4 - 10^6\,M_{\odot}$, respectively.
It is essential to investigate the properties and formation/evolution processes of GMCs because they are known to be major sites of high-mass star formation, which eventually drives the evolution of galaxies.

So far, a lot of studies have investigated various GMC properties and their relationships.
In the MW, \cite{larson1981} found that the internal velocity dispersions of the molecular clouds are well correlated with their sizes and masses, and also reported that these correlations (i.e., scaling relations) can be expressed as the power-law form.
\cite{solomon1987} measured the velocity dispersions, sizes, virial masses, and CO luminosities for 273 GMCs in the Galactic disk, and found that the velocity dispersion is proportional to the 0.5 power of the size.
They also found a tight relationship, over four orders of magnitude, between the virial mass and the CO luminosity with a power-law slope of $\sim$0.8.

Such GMC studies were expanded to the Local Group galaxies outside the MW.
\cite{fukui2008} performed a CO survey toward the Large Magellanic Cloud (LMC) at a spatial resolution of $\sim$40\,pc, and identified 272 GMCs with a mass range from $2 \times 10^4\,M_{\odot}$ to $7 \times 10^6\,M_{\odot}$ \citep[see also][]{fukui1999}.
In addition, \cite{kawamura2009} examined spatial comparisons of these GMCs with young star clusters (YSCs) and H\,{\sc ii} regions and found that the GMCs can be classified into three types:
(1) GMCs associated with no H\,{\sc ii} regions nor YSCs, (2) GMCs associated only with small H\,{\sc ii} regions, but with no YSCs, and (3) GMCs associated with both YSCs and large H\,{\sc ii} regions.
Such a classification of GMCs according to the activities of high-mass star formation likely reflects their evolutionary sequence.
In addition, GMC surveys have been often conducted toward M33, which is one of the nearest spiral galaxies \citep[e.g.,][]{engargiola2003, rosolowsky2007, gratier2012, miura2012, corbelli2017}.
These studies identified more than 100 GMCs \citep[in particular, more than 500 GMCs by][]{corbelli2017} over the M33 disk at $\sim$50\,pc resolution, and discussed timescales and the evolutionary stages of GMCs based on the comparison with H\,{\sc ii} regions and YSCs as well as the LMC studies.

High-angular resolution observations by millimeter-wave interferometers enabled to perform the unbiased GMC surveys even toward external spiral galaxies.
\cite{colombo2014} reported the GMC catalog, which contains $\sim$1500 individual objects in the grand-design spiral galaxy M51 at $\sim$40\,pc resolution using data from the PdBI Arcsecond Whirlpool Survey \citep{schinnerer2013}.
They proposed that large-scale dynamical processes and feedback from high-mass star formation cause environmental variations in the GMC properties and mass distributions, and also suggested that $\sim$30\% of GMCs in M51 are unbound.
More recently, PHANGS-ALMA survey mapped CO($J=2-1$) line emission at $\sim$1$\arcsec$ resolution toward 90 nearby star-forming galaxies \citep{leroy2021}.
In particular, \cite{rosolowsky2021} identified 4986 molecular clouds at a common 90\,pc resolution and measured their properties for ten subsamples.
They found that the physical properties of clouds vary among galaxies, both as a function of galactocentric radius and as a function of the dynamical environment (e.g., bar, spiral arm, and inter-arm).

However, these earlier studies for external spiral galaxies are likely biased toward the massive ($\geq 10^5\,M_{\odot}$) population of molecular clouds
except for the case of M33 \citep[e.g., a small GMC down to $2.4 \times 10^4 M_{\odot}$ is recovered by][]{corbelli2017}.
To understand the complex hierarchical structures of molecular gas and also to understand the evolution of molecular clouds in galaxies,
smaller and less massive ($< 10^5\,M_{\odot}$) molecular clouds should be investigated \citep[e.g., the slope of molecular cloud mass functions changes with evolution processes;][]{kobayashi2017, kobayashi2018}.
Thus, we need further molecular-cloud surveys covering such less massive clouds in nearby galaxies as a complementary study to PHANGS-ALMA survey.

In this paper, we present the results of a new CO($J=2-1$) survey toward almost the whole molecular-gas disk of M33 conducted with the Atacama Compact Array (ACA) stand-alone mode of ALMA.
The distance to M33 is estimated to be 840\,kpc \citep{freedman1991, galleti2004}; thus, 1$\arcsec$ corresponds to 4\,pc.
The inclination of M33 is 55$^{\circ}$ \citep{koch2018}.
Its proximity and relatively small inclination angle have enabled many researchers to study the ISM and high-mass star formation over the wide area of the M33 disk at a few $\times$ 10\,pc scale
\citep{engargiola2003, rosolowsky2007, onodera2010, gratier2010, tosaki2011, gratier2012, miura2012, onodera2012, miura2014, druard2014, gratier2017, corbelli2017}.
In addition, recent studies based on ALMA 12-m array observations revealed complicated internal molecular-gas structures within some especially massive ($\sim 10^6\,M_{\odot}$) GMCs of M33 at 1\,--\,2\,pc scale \citep[e.g.,][]{tokuda2020, muraoka2020, kondo2021, sano2021}.
Thus, M33 is a unique target to investigate the hierarchical structure of molecular gas in face-on spiral galaxies from parsec to kiloparsec scales.
The basic properties of M33 are summarized in Table~\ref{tab:m33}.
The main purposes of the new ACA observations are to obtain the spatial distribution in CO($J=2-1$) emission with the higher sensitivity and higher angular resolution compared to earlier studies in M33
and to identify low-mass ($< 10^5\,M_{\odot}$) clouds as well as high-mass ($\geq 10^5\,M_{\odot}$) clouds.
This surely becomes an important step to understand the hierarchical structures of molecular gas and the evolution of molecular clouds in galaxies.

The structure of this paper is as follows.
In Section~\ref{sec:obs}, we describe the detail of the ACA observations and data reduction.
Then, we present the overall molecular-gas structures in CO($J=2-1$) emission at $\sim$30\,pc resolution in M33 in Section~\ref{sec:res}.
In Section~\ref{sec:clouds}, we describe the procedure of cloud decomposition based on $^{12}$CO($J=2-1$) cube data, and summarize the basic properties of cataloged molecular clouds.
In Section~\ref{sec:scaling}, we examine the scaling relations for the molecular clouds in M33.
We compare the cataloged molecular clouds in this study with the earlier GMC catalog in M33 summarized by \cite{corbelli2017} in Section~\ref{sec:comparison}.
Finally, we discuss the relationship between the properties of molecular clouds and the high-mass star formation in M33 in Section~\ref{sec:sf}.

\begin{deluxetable*}{llc}
\tabletypesize{\scriptsize}
\tablewidth{0pt} 
\tablenum{1}
\tablecaption{General properties of M33
\label{tab:m33}}
\tablehead{
\colhead{Parameter}         & \colhead{Value}                        & \colhead{Reference}
}                                                                                                                                                                                                                                                                                                                  
\startdata                                                                                                                                                                                                                                                                                                         
IR center (J2000):          &                                        & (1)      \\
\,\,\,\, Right Ascension    & 1$^{\rm h}$33$^{\rm m}$50$^{\rm s}$.9  &          \\
\,\,\,\, Declination        & 30$^{\circ}$39\arcmin37\arcsec         &          \\
Distance                    & 840\,kpc                               & (2), (3) \\
LSR velocity                & 170\,km\,s$^{-1}$                      & (4)      \\
Inclination                 & 55$^{\circ}$                           & (5)      \\
Position angle              & 21$^{\circ}$                           & (5)      \\
Stellar mass                & $4.8 \times 10^9$\,$M_{\odot}$         & (6)      \\
Molecular gas mass          & $3.1 \times 10^8$\,$M_{\odot}$         & (4)
\enddata
\tablecomments{References. (1) \cite{skrutskie2006}, (2) \cite{freedman1991}, (3) \cite{galleti2004}, (4) \cite{druard2014}, (5) \cite{koch2018}, (6) \cite{corbelli2014}
}
\end{deluxetable*}

\section{Observations and Data Reduction} \label{sec:obs}

\begin{figure*}[ht!]
\epsscale{0.55}
\plotone{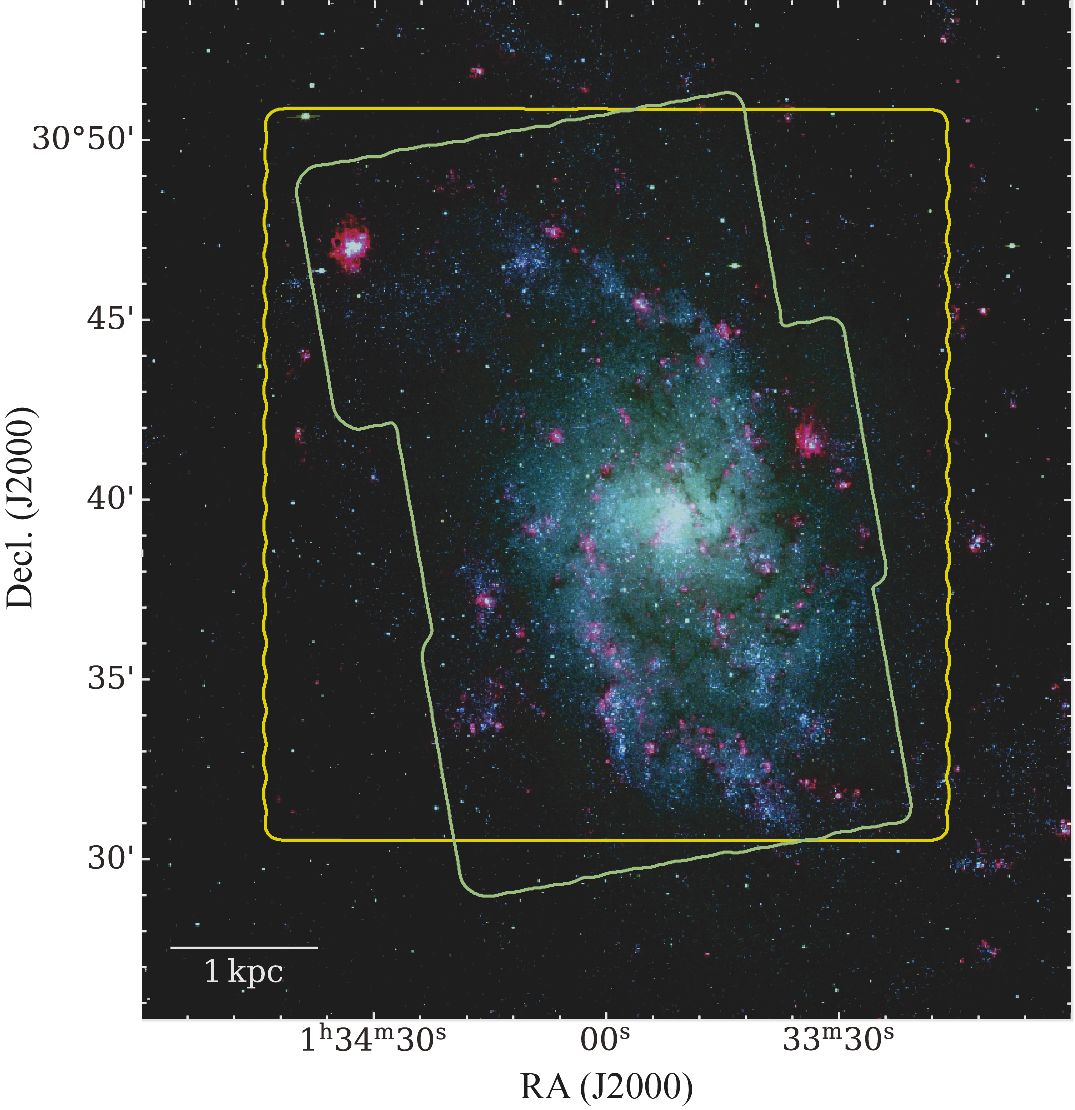}
\caption{Three-color composite image constructed from the B-band (blue), V-band (green), and H$\alpha$ (red) images of M33 taken with the Mayall 4\,m telescope \citep[][]{massey2006, massey2007}. 
The yellow line indicates the observed field of the project code 2018.A.00058.S, and the green line indicates that of 2017.1.00901.S and 2019.1.01182.S.
}\label{fig:obsarea}
\end{figure*}

Observations toward M33 were carried out in Band~6 (211\,--\,275\,GHz) with the ACA 7\,m antennas between 2019 August and 2021 August (project code 2018.A.00058.S).
The target molecular lines were $^{12}$CO($J=2-1$), $^{13}$CO($J=2-1$), and C$^{18}$O($J=2-1$).
The bandwidths of the correlator settings were 117.19\,MHz with 1920\,channels for the $^{12}$CO line and 960\,channels for $^{13}$CO and C$^{18}$O lines.
The target field was the rectangle with the size of 1100$\arcsec$ $\times$ 1180$\arcsec$ (4.5\,kpc $\times$ 4.8\,kpc), covering most of the molecular-gas disk of M33.
The total number of mosaic fields is 3129.
In addition to this, we retrieved the ALMA archival data (project code 2017.1.00901.S and 2019.1.01182.S), which also observed the molecular-gas disk of M33 by ACA 7\,m antennas with almost the same spectral settings as our observations.
Prior to the imaging process, we concatenated all visibilities obtained in each science goal with a total number of 36.
This data reduction strategy is the same as the previously published large-scale ACA mapping project on the Small Magellanic Cloud (SMC) \citep{tokuda2021}.
Figure~\ref{fig:obsarea} shows the eventually observed field.

We used Common Astronomy Software Application (CASA) package \citep{mcmullin2007} version 5.4.0 in the data reduction.
We applied the standard calibration scheme provided by the ALMA observatory while we performed the imaging process.
We used the \texttt{tclean} task with \texttt{multi-scale} deconvolver \citep{kepley2020} to recover extended emission as much as possible.
In \texttt{tclean} task, we applied the natural weighting and used the \texttt{auto-multithresh} procedure to identify automatically regions containing emission in the dirty and residual images.
We continued the deconvolution process until the intensity of the residual image attains the $\sim$1\,$\sigma$ noise level.
The beam size and the rms noise level for each emission are summarized in Table~\ref{tab:emission}.

To evaluate the missing flux of the ACA observations, we measured the global $^{12}$CO($J=2-1$) luminosities over the M33 disk obtained by the ACA 7\,m antennas and by the IRAM 30\,m telescope \citep{druard2014}.
We found the global $^{12}$CO($J=2-1$) luminosity with ACA 7\,m antennas $L_{\rm CO}^{\rm ACA} = 7.6 \times 10^6$\,K\,km\,s$^{-1}$\,pc$^2$ over the observed region,
and that with the IRAM 30\,m telescope $L_{\rm CO}^{\rm IRAM} = 2.1 \times 10^7$\,K\,km\,s$^{-1}$\,pc$^2$ for the same area.
This indicates the global missing flux of $^{12}$CO($J=2-1$) emission of $60-70$\%, which mainly corresponds to diffuse components of molecular gas.
To compensate for such diffuse components, we combined the ACA 7\,m-array $^{12}$CO($J=2-1$) data with the IRAM 30\,m data using the \texttt{feather} task.
Hereafter, we refer to the pre-combined ACA 7\,m-array $^{12}$CO($J=2-1$) data as ``stand-alone ACA $^{12}$CO($J=2-1$)'' data, and to the combined $^{12}$CO($J=2-1$) data as ``ACA+IRAM $^{12}$CO($J=2-1$)'' data.
The beam size and the rms noise level of ACA+IRAM $^{12}$CO($J=2-1$) data are the same as those of the stand-alone ACA $^{12}$CO($J=2-1$) data.

\begin{deluxetable*}{llcc}
\tabletypesize{\scriptsize}
\tablewidth{0pt} 
\tablenum{2}
\tablecaption{Properties of each line emission
\label{tab:emission}}
\tablehead{
\colhead{Line}      & \colhead{Beam Size}                                     & \colhead{Rms Noise Level} & \colhead{Velocity Resolution} \\
}                                                                                                                                                                                                                                                                                                                 
\startdata                                                                                                                                                                                                                                                                                                         
$^{12}$CO($J=2-1$)  & 7\farcs31 $\times$ 6\farcs50 (30\,pc $\times$ 26\,pc)   & 39\,mK   & 0.7\,km\,s$^{-1}$ \\
$^{13}$CO($J=2-1$)  & 7\farcs72 $\times$ 6\farcs86 (31\,pc $\times$ 27\,pc)   & 30\,mK   & 1.4\,km\,s$^{-1}$ \\
C$^{18}$O($J=2-1$)  & 7\farcs82 $\times$ 6\farcs96 (31\,pc $\times$ 28\,pc)   & 22\,mK   & 1.6\,km\,s$^{-1}$ 
\enddata
\end{deluxetable*}

\section{CO Maps} \label{sec:res}

From the reduced three-dimensional cube data, we examine the zeroth moment (i.e., velocity-integrated intensity) in $^{12}$CO($J=2-1$) and $^{13}$CO($J=2-1$) emission.
To minimize the effect of the noise, we determined the velocity channel in which the CO emission is expected to appear using the atomic (H\,{\sc i}) data \citep{koch2018} as follows.
Firstly, we convolved the H\,{\sc i} data whose original angular resolution is 20$\arcsec$ to 40$\arcsec$ in order to reduce the effect of the anomalous H\,{\sc i} velocity components.
Then, we regridded them to match our CO data and determined the representative H\,{\sc i} velocity $V_{\rm rep}$ in each pixel.
Finally, we calculated the zeroth moment in $^{12}$CO($J=2-1$) emission from $V_{\rm rep}$\,$-$\,30\,km\,s$^{-1}$ to $V_{\rm rep}$\,$+$\,30\,km\,s$^{-1}$.
Although \citet{rosolowsky2007} reported that 90\% of the velocity separation between CO and H\,{\sc i} is within 20\,km\,s$^{-1}$, each CO line typically has a velocity width of 5\,--\,10\,km\,s$^{-1}$.
In fact, we found that some molecular clouds dropped $\sim$30\% of CO flux if we apply the velocity range of $V_{\rm rep}$\,$\pm$\,20\,km\,s$^{-1}$.
To correctly measure the CO intensity in M33, we needed the velocity range of $V_{\rm rep}$\,$\pm$\,30\,km\,s$^{-1}$ for the calculation of the $^{12}$CO($J=2-1$) zeroth moment.
We also calculated the $^{13}$CO($J=2-1$) zeroth moment from $V_{\rm rep}$\,$-$\,30\,km\,s$^{-1}$ to $V_{\rm rep}$\,$+$\,30\,km\,s$^{-1}$.

\begin{figure*}[ht!]
\epsscale{1.15}
\plotone{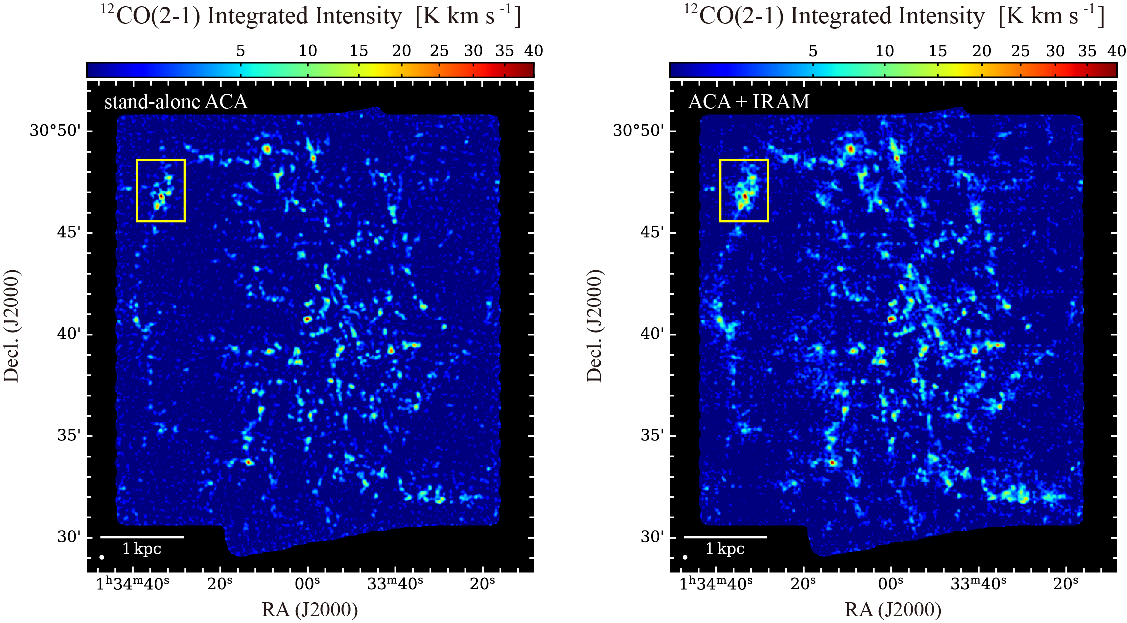}
\caption{Integrated intensity maps in $^{12}$CO($J=2-1$) from the stand-alone ACA data (left) and the ACA+IRAM data (right).
The yellow rectangle indicates the NGC\,604 region. A zoomed-in view of this region is shown in Figure~\ref{fig:zoomin12CO}.
The synthesized beam is shown in the lower left corner.
}\label{fig:12COmap}
\end{figure*}

\begin{figure*}[ht!]
\epsscale{1.15}
\plotone{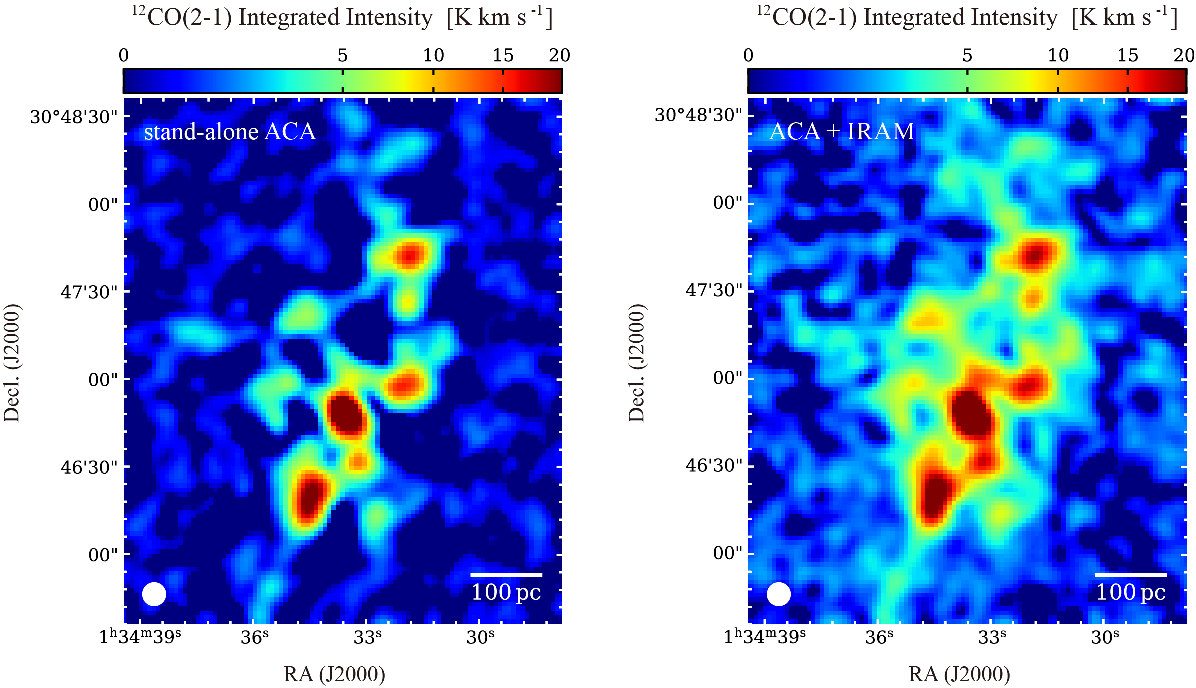}
\caption{Zoomed-in view of the NGC\,604 region in Figure~\ref{fig:12COmap}.
Diffuse molecular-gas components between bright GMCs are well recovered.
The synthesized beam is shown in the lower left corner.
}\label{fig:zoomin12CO}
\end{figure*}

Figure~\ref{fig:12COmap} shows the integrated intensity maps in $^{12}$CO($J=2-1$) from the stand-alone ACA data and the ACA+IRAM data, respectively.
These $^{12}$CO($J=2-1$) maps clearly depict the molecular-gas structure within M33 at 30\,pc resolution.
We can easily find a lot of individual molecular clouds over the M33 disk.
The ACA+IRAM $^{12}$CO($J=2-1$) map properly recovers diffuse components of molecular gas, which are missed in the stand-alone ACA map.
We show an evident case, $^{12}$CO($J=2-1$) integrated intensity map for GMCs associated with the giant H\,{\sc ii} region NGC\,604, in Figure~\ref{fig:zoomin12CO}.

\begin{figure*}[ht!]
\epsscale{1.15}
\plotone{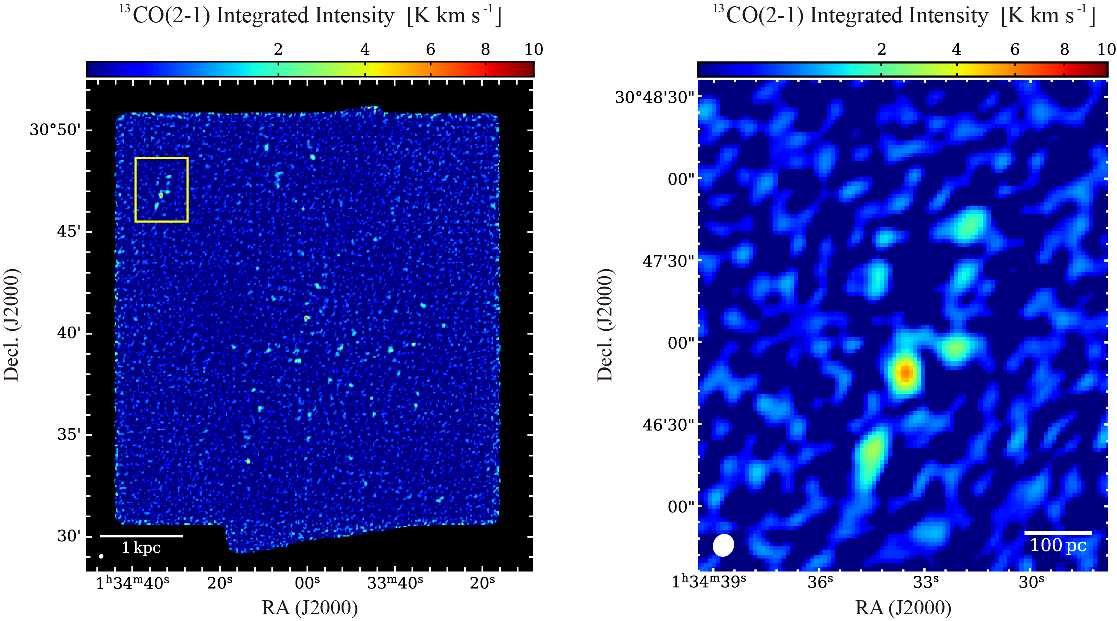}
\caption{Integrated intensity map in $^{13}$CO($J=2-1$) emission of M33 obtained by the ACA 7\,m-array (left) and its zoomed-in view of the NGC\,604 region (right). 
The synthesized beam is shown in the lower left corner.
}\label{fig:13COmap}
\end{figure*}

The integrated intensity map in $^{13}$CO($J=2-1$) emission over the M33 disk is shown in the left panel of Figure~\ref{fig:13COmap}.
A lot of $^{13}$CO($J=2-1$) sources are detected.
They correspond to moderately dense gas whose density is $\gtrsim$10$^{3}$\,cm$^{-3}$ within the $^{12}$CO cloud.
The zoomed-in view of the NGC\,604 region is shown in the right panel of Figure~\ref{fig:13COmap}.

Note that we found no significant C$^{18}$O($J=2-1$) emission in the ACA map.
The rms noise level of 22\,mK yields a 3\,$\sigma$ upper limit of 66\,mK.
To check the validity of the upper limit, we retrieved the ALMA archival data (project code 2017.1.00461.S) and examined the C$^{18}$O($J=2-1$) emission in a GMC associated with NGC\,604.
We found that the peak temperature of the strongest C$^{18}$O($J=2-1$) emission is $\sim 1$\,K at an angular resolution of 0\farcs3 (1.2\,pc) and its spatial extent is less than 1$\arcsec$.
Then, we convolved the C$^{18}$O($J=2-1$) emission to 7\farcs5 and found that the peak temperature decreases down to $\sim 30$\,mK, which corresponds to 1.4\,$\sigma$ in the ACA C$^{18}$O($J=2-1$) map.
Thus, we consider that the beam smearing effect makes C$^{18}$O($J=2-1$) emission undetectable in the ACA 30\,pc-resolution map.

\section{Cloud Decomposition} \label{sec:clouds}

As shown in Figure~\ref{fig:12COmap}, the structure of molecular clouds is highly complex and hierarchical over the M33 disk.
To identify individual emission structures in an objective way and to investigate the properties of molecular clouds, 
we employed \texttt{PYCPROPS} \citep{rosolowsky2021}, a Python implementation of the algorithm to catalog molecular clouds, \texttt{CPROPS} \citep{rosolowsky2006}.
We use the ACA+IRAM $^{12}$CO($J=2-1$) data in the following analyses.

Firstly, we convolved the ACA+IRAM $^{12}$CO($J=2-1$) cube data to 7\farcs5 in order to identify molecular clouds by a circular beam.
Then, we made masked cube data using \texttt{cprops\_mask} provided by \cite{rosolowsky2021}. 
The criteria of the emission mask are that high significance emission is required to be more than 4\,$\sigma$ in continuous three velocity channels and low significance emission
which is adjacent to high significance emission is required to be more than 3\,$\sigma$ in continuous three velocity channels at least over the size of the ACA 7\farcs5 beam.
Although the default settings of \texttt{cprops\_mask} are 4\,$\sigma$ and 2\,$\sigma$ for high and low significance masks, respectively,
we found that the emission masks with these default settings are not suitable for the ACA+IRAM $^{12}$CO($J=2-1$) cube data.
In particular, the low significance mask does not reject fake emission (i.e., noise) at the cloud edge.
Thus, we carefully tuned the rms thresholds, and finally we adopted 3\,$\sigma$ for low significance masks.

\texttt{PYCPROPS} firstly searches for all local maxima in the emission-masked cube data and measures the peak temperature in each local maximum, $T_{\rm max}$.
When a local maximum has at least one other neighbor whose peak temperature is $T_{\rm merge}$, \texttt{PYCPROPS} compares $T_{\rm max}$ and $T_{\rm merge}$.
The neighbor is rejected if $T_{\rm max} - T_{\rm merge}$ is less than 2\,$\sigma$, which means that such a local maximum is likely a noise fluctuation.
The criterion of 2\,$\sigma$ is a default value recommended by \cite{rosolowsky2008}.
Then, \texttt{PYCPROPS} determines if the spatial and spectral separations between local maxima are adequate or not.
We adopted the beam size (7\farcs5) as a minimum spatial separation and also adopted 7\,km s$^{-1}$ as a minimum spectral separation,
which corresponds to a typical velocity width of a GMC with the size of $\sim$30\,pc considering the Galactic size-linewidth relation \citep{solomon1987}.
If either spatial separation or spectral separation between local maxima does not satisfy the above threshold, the local maximum which has a smaller $T_{\rm max}$ is rejected.
Through these processes, \texttt{PYCPROPS} identifies a set of significant local maxima.
We treat these local maxima as seeds to assign all the emission to molecular clouds.
To do this, we use a watershed algorithm, which associates all the cube pixels in the emission-masked data with a local maximum.
Some pixels are already assigned to a single local maximum, while the remainder (including rejected local maxima in the above processes) are assigned to any of the local maxima by the watershed algorithm.
More details on the \texttt{PYCPROPS} algorithm are summarized in \cite{rosolowsky2021}.
Finally, \texttt{PYCPROPS} identified 886 molecular clouds.

\begin{figure*}[ht!]
\epsscale{1.15}
\plotone{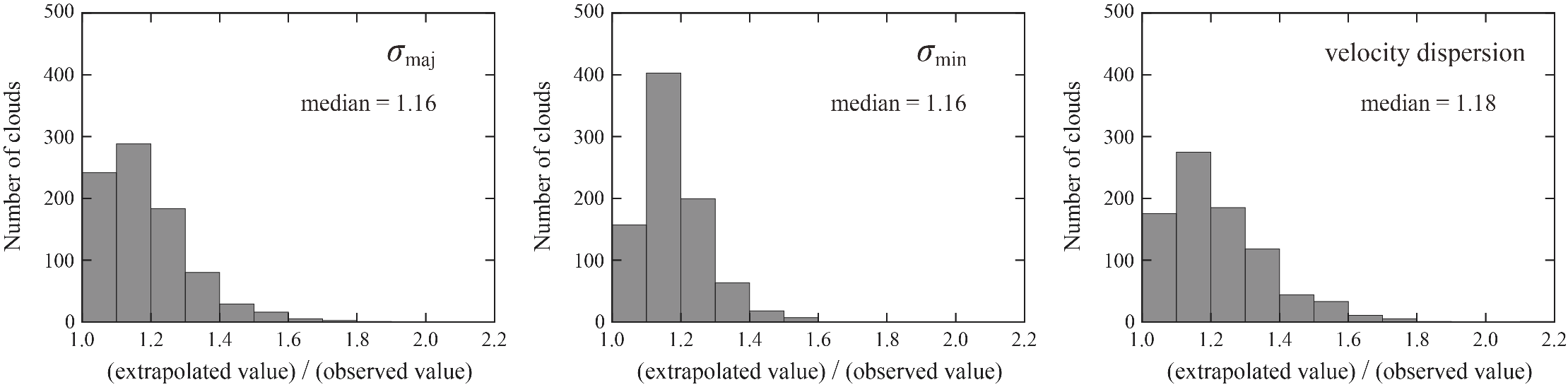}
\caption{Frequency distributions of the ratio between the extrapolated value and the observed value for the second moment of the major axis ($\sigma_{\rm maj}$; left),
that of the minor axis ($\sigma_{\rm min}$; center), and velocity dispersion (right), respectively.
}\label{fig:extrapolation}
\end{figure*}

\texttt{PYCPROPS} gives the basic properties of the identified clouds, including the extrapolated 2nd moment of the emission along the major and minor axes $\sigma_{\rm maj}$ and $\sigma_{\rm min}$ in parsec,
the position angle of the major axis $\phi$, the extrapolated velocity dispersion $\sigma_{v, {\rm ext}}$, and the integrated $^{12}$CO($J=2-1$) flux $S$ within each cloud.
The extrapolated cloud properties are calculated to reduce observational bias \citep{rosolowsky2006}.
In Figure~\ref{fig:extrapolation}, we showed frequency distributions of the ratio between the extrapolated value and the observed value for cloud size and velocity dispersion.
The extrapolated value is typically 10\,--\,20\% larger than the observed value.

We calculated the intrinsic spherical radius $R$ by the deconvolution of the ACA+IRAM beam, $\sigma_{\rm beam}$, as follows:

\begin{eqnarray}
R = 1.91 \sqrt{(\sigma_{\rm maj}^2 - \sigma_{\rm beam}^2)^{0.5} (\sigma_{\rm min}^2 - \sigma_{\rm beam}^2)^{0.5}}.
\end{eqnarray}
Here, $\sigma_{\rm beam}$ is calculated as $7\farcs5 \times 4$ = 30, where the factor of 4 is the spatial size in parsec of 1$\arcsec$ at the distance of M33 (840\,kpc).
The coefficient 1.91 converts the rms size to the effective spherical radius of the cloud \citep[e.g.,][]{solomon1987, rosolowsky2006}. 
We treat $R$ as the intrinsic spherical radius of each molecular cloud.
Note that if $\sigma_{\rm min}$ is smaller than $\sigma_{\rm beam}$, the resultant $R$ is not properly defined.
We do not consider such small clouds further in this paper.

We also deconvolved the velocity dispersion $\sigma_{v, {\rm ext}}$ as follows:

\begin{eqnarray}
\sigma_v = \sqrt{\sigma_{v, {\rm ext}}^2 - \frac{\sigma_{v, {\rm chan}}^2}{2 \pi}},
\end{eqnarray}
where $\sigma_{v, {\rm chan}}$ is the velocity resolution element, which is related to the velocity channel width ($\Delta V_{\rm chan}$ = 0.7\,km\,s$^{-1}$) as $\sigma_{v, {\rm chan}} = \Delta V_{\rm chan}/(2 \sqrt{2\,{\rm ln}\,2})$.

\begin{figure*}[ht!]
\epsscale{1.15}
\plotone{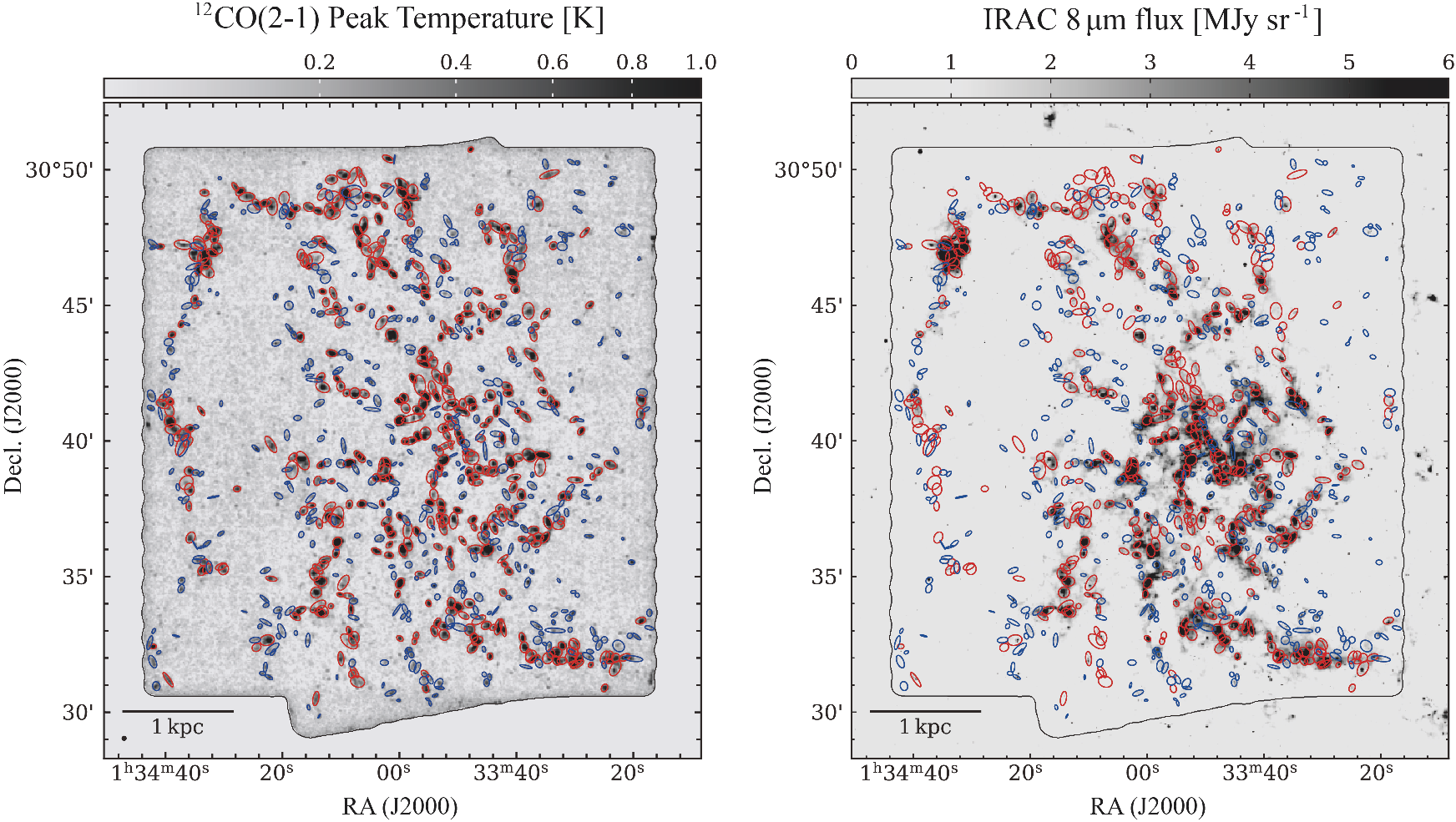}
\caption{Distribution of 848 molecular clouds in M33 superposed on the ACA+IRAM combined $^{12}$CO($J=2-1$) peak temperature map (left) and the map of $Spitzer$/IRAC $8\,\mu$m flux obtained by \citet{dale2009}(right).
The molecular clouds are represented as ellipses, whose sizes and orientation indicate the extrapolated and deconvolved major and minor axes, and the measured position angle.
Red ellipses represent high-mass clouds ($M_{\rm CO} \geq 10^5\,M_{\odot}$) and blue ellipses indicate low-mass clouds ($M_{\rm CO} < 10^5\,M_{\odot}$).
The black line indicates the eventually observed field by ACA.
}\label{fig:cloudsmap}
\end{figure*}

We do not consider molecular clouds lying at the edge of the ACA field-of-view (FOV) further because the primary beam correction causes larger uncertainties in the obtained properties of molecular clouds.
Thus, we consider 848 molecular clouds after excluding small clouds with undefined radius and clouds at the edge of ACA FOV from the originally identified clouds by \texttt{PYCPROPS}.
Figure~\ref{fig:cloudsmap} shows the spatial distribution of 848 molecular clouds, whose deconvolved sizes and the measured position angle are represented, in the M33 disk.

\subsection{Luminosities and Masses} \label{sec:masslumi}

From the basic properties of molecular clouds, we calculated additional properties such as the CO luminosity $L_{\rm CO} = SD^2$ where $D = 840$\,kpc and cloud masses.
The luminosity-based mass (for $^{12}$CO, hereafter $M_{\rm CO}$) which includes the helium contribution is calculated as

\begin{eqnarray}
\frac{M_{\rm CO}}{M_{\odot}} = 4.35 \, \frac{X_{\rm CO}}{2.0 \times 10^{20}\,{\rm cm}^{-2} ({\rm K\,km\,s}^{-1})^{-1}} \, \frac{L_{\rm CO}}{{\rm K\,km\,s}^{-1}\,{\rm pc}^2} \, R_{21}^{-1},
\end{eqnarray}
where $X_{\rm CO}$ is CO-to-H$_2$ conversion factor and $R_{21}$ is the $^{12}$CO($J=2-1$)/$^{12}$CO($J=1-0$) intensity ratio.
We adopted a constant $X_{\rm CO}$ of $4.0 \times 10^{20}$\,cm$^{-2}$ (K\,km\,s$^{-1}$)$^{-1}$ \citep{gratier2017} over the M33 disk.
To determine the appropriate $R_{21}$ value in this study, we examined the pre-existing single-dish measurements of $^{12}$CO in M33 for $J=1-0$ \citep{tosaki2011} and $J=2-1$ \citep[][]{gratier2010, druard2014} transitions.
We found that the average $R_{21}$ in M33 is 0.60, but this value is lower than the previously-reported $R_{21}$ of 0.8 \citep{druard2014}.
In the MW, the reported $R_{21}$ is 0.64 \citep{yoda2010}.
In addition, recent studies reported that the mean of $R_{21}$ in nearby galaxies is 0.6\,--\,0.7 \citep[e.g.,][]{yajima2021, denbrok2021, leroy2022}.
These $R_{21}$ values are consistent with the newly-obtained one in M33, 0.60.
Thus, we adopted a constant $R_{2-1/1-0}$ of 0.60 across the M33 disk in this study.
Note that, as pointed out by \citet{yajima2021}, $R_{21}$ varies within an individual galaxy; in fact, $R_{21}$ in M33 varies from position to position, typically ranging from 0.4 to 0.8.
Therefore we consider that the assumption of a constant $R_{2-1/1-0}$ over the M33 disk yields an error of about 30\%.
From the emission masks and \texttt{PYCPROPS} parameters, the detection limit of $M_{\rm CO}$ is calculated to be $3 \times 10^3\,M_{\odot}$,
while the actual lowest mass of the molecular clouds is $7 \times 10^3\,M_{\odot}$.

In Figure~\ref{fig:cloudsmap}, each molecular cloud is color-coded according to its $M_{\rm CO}$,
i.e., red ellipses represent high-mass clouds ($M_{\rm CO} \geq 10^5\,M_{\odot}$) and blue ellipses indicate low-mass clouds ($M_{\rm CO} < 10^5\,M_{\odot}$).
In addition, the spatial comparison between molecular clouds and $Spitzer$/IRAC $8\,\mu$m emission \citep{dale2009} is displayed.
In the spiral arm region, many high-mass clouds are associated with the strong (typically $> 2$\,MJy\,sr$^{-1}$) 8\,$\mu$m emission, which likely traces high-mass star-forming regions \citep[e.g.,][]{calzetti2005, wu2005, calzetti2007, crocker2013}.
On the other hand, low-mass clouds tend to be apart from such $8\,\mu$m-bright sources and to exist in the inter-arm region.

We examine the mass fraction of the molecular clouds to the total molecular gas over the ACA-observed area.
The total mass of the molecular clouds is derived to be $1.6 \times 10^8\,M_{\odot}$ by summing up their $M_{\rm CO}$ values.
We calculated the global ACA+IRAM $^{12}$CO($J=2-1$) of $2.0 \times 10^7$\,K\,km\,s$^{-1}$\,pc$^2$, which yields the total molecular gas mass of $2.9 \times 10^8\,M_{\odot}$\footnote{The $^{12}$CO($J=2-1$) data obtained by IRAM 30\,m telescope \citep{druard2014} gives its luminosity of $2.1 \times 10^7$\,K\,km\,s$^{-1}$\,pc$^2$ within the ACA FOV (see Section~\ref{sec:obs}). If we assume $X_{\rm CO}$ of $4.0 \times 10^{20}$\,cm$^{-2}$ (K\,km\,s$^{-1}$)$^{-1}$ and $R_{21}$ of 0.6, this luminosity yields the total molecular gas mass of $3.0 \times 10^8\,M_{\odot}$, which is well consistent with the molecular gas mass of $2.9 \times 10^8\,M_{\odot}$ derived from ACA+IRAM $^{12}$CO($J=2-1$) data.}.
Thus, the mass fraction of molecular clouds to the total molecular gas is 55\%.
This is similar to the case in M51; \cite{pety2013} reported that about half of the CO luminosity arises from molecular clouds and the other half from diffuse components of molecular gas.

We also calculated the virial mass as $M_{\rm Vir} = 1040\,R\,\sigma_v^2$ for a spherical and virialized cloud with a density profile of $\rho \propto r^{-1}$ \citep{bolatto2013}.
Its relationship with $M_{\rm CO}$ is discussed in Section~\ref{sec:scaling}.

\subsection{$^{13}$CO($J=2-1$) Emission} \label{sec:13CO}

We examined $^{13}$CO($J=2-1$) emission for each cloud.
The criteria for the ``detection'' of $^{13}$CO($J=2-1$) emission are as follows.
Firstly, we drew the $^{12}$CO($J=2-1$) spectrum at the $^{12}$CO($J=2-1$) peak of each cloud
and defined the ``line channels'', which are successive velocity channels where significant $^{12}$CO($J=2-1$) emission exists.
Then, we examined the $^{13}$CO($J=2-1$) spectrum within the line channels.
If $^{13}$CO($J=2-1$) emission exceeds 4\,$\sigma$ for successive 2\,channels or exceeds 3\,$\sigma$ for successive 3\,channels,
we treat the $^{13}$CO($J=2-1$) emission as temporarily-detected.
In addition, we calculated the $^{13}$CO($J=2-1$) integrated intensity within the line channels, and derive its signal-to-noise (S/N) ratio.
If the S/N ratio of the temporarily-detected $^{13}$CO($J=2-1$) intensity exceeds 3, we finally treat the $^{13}$CO($J=2-1$) emission as significantly detected.
We confirmed significant $^{13}$CO($J=2-1$) emission for 173 clouds, and thus the resultant $^{13}$CO($J=2-1$) detection rate is 20\%.

\begin{figure*}[ht!]
\epsscale{0.5}
\plotone{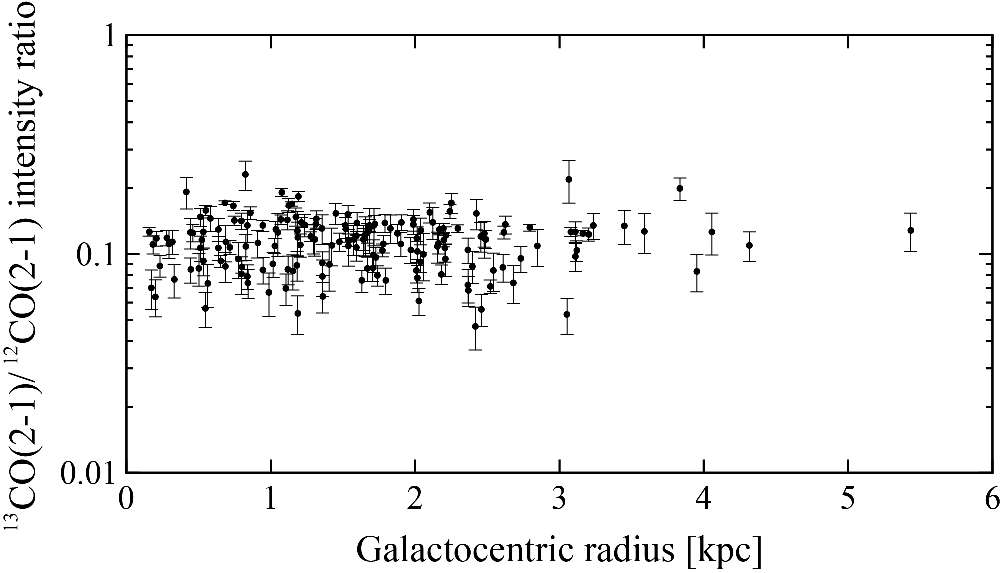}
\caption{$^{13}$CO($J=2-1$)/$^{12}$CO($J=2-1$) intensity ratio ($R_{13/12}$) as a function of the galactocentric radius for the $^{13}$CO($J=2-1$) detected clouds in M33.
}\label{fig:distratio}
\end{figure*}

We examined the $^{13}$CO($J=2-1$)/$^{12}$CO($J=2-1$) intensity ratio (hereafter $R_{13/12}$) for the $^{13}$CO($J=2-1$) detected clouds.
We found that $R_{13/12}$ in M33 is almost constant on the galactocentric radius as shown in Figure~\ref{fig:distratio},
and the typical $R_{13/12}$ is $\sim0.1$.
This value is similar to that in the disk of M51 \citep{denbrok2022}, and also similar to that for $J=1-0$ transition (i.e., $^{13}$CO($J=1-0$)/$^{12}$CO($J=1-0$) ratio)
measured in nearby galaxy disks \citep[e.g.,][]{paglione2001, hirota2010, watanabe2011, muraoka2016, cao2017, cormier2018, yajima2019, topal2020, morokuma2020, cao2023}.

\subsection{Catalog Description} \label{sec:catalog}

We summarized the properties of 848 clouds in M33 as a catalog, which do not include small clouds with undefined radius and clouds at the edge of ACA FOV.
We assigned the ID number of the clouds in order of increasing the galactocentric radius.
Table~\ref{tab:clouds} presents the first 10 and last 10 clouds of the catalog, and the full version is available online.

The uncertainty of each property was evaluated using a bootstrapping method implemented in \texttt{PYCPROPS}.
Considering that the cloud consists of $N$ data points, we generated a trial cloud by $N$ times random sampling of the data allowing the same data to be sampled more than once.
Then, we measured the properties of the trial cloud.
We repeated the resampling and remeasuring 10,000 times for each cloud, and evaluated the uncertainties.
The final uncertainty in each property is the median absolute deviation of the bootstrapped values scaled up by the square root of the over-sampling rate, which corresponds to the number of pixels per beam size.
This scaling accounts for the fact that pixels within the same beam are not independent (i.e., correlated with each other).

In this molecular-cloud catalog, we noted the S/N ratio of $^{12}$CO($J=2-1$) brightness temperature at the CO peak position in each GMC.
As reported in \cite{rosolowsky2006}, the \texttt{CPROPS} algorithm requires a minimum S/N ratio of 10 for stable recovery of cloud properties.
Since our catalog includes 147 molecular clouds whose S/N ratio is lower than 10, we examine the properties of such low-S/N clouds in the following analyses.
The minimum S/N ratio of the cataloged cloud is 6.1. In addition, we checked a GMC counterpart identified with the IRAM 30\,m telescope \citep{corbelli2017}.
We make a comparison between the two catalogs in Section~\ref{sec:comparison}.

\begin{longrotatetable}
\begin{deluxetable*}{cccccccccccccccc}
\tabletypesize{\scriptsize}
\tablewidth{0pt} 
\tablenum{3}
\tablecaption{List of cataloged molecular clouds
\label{tab:clouds}}
\tablehead{
\colhead{ID}  & \colhead{R.A.} & \colhead{Decl.} & \colhead{$R_{\rm gal}$} & \colhead{$V_{\rm LSR}$} & \colhead{$R$}   & \colhead{$\sigma_v$} & \colhead{$M_{\rm CO}$}  & \colhead{$M_{\rm Vir}$} & \colhead{$I_{12 {\rm CO}}$} & \colhead{$^{12}$CO S/N} 		 & \colhead{$I_{13 {\rm CO}}$} & \colhead{$R_{13/12}$} & $\phi$            & $b/a$          & IRAM ID        \\[-1.0mm]
              & (deg.)         & (deg.)          & (kpc)                   & (km\,s$^{-1}$)          & (pc)            & (km\,s$^{-1}$)       & ($10^4\,M_{\odot}$)     & ($10^4\,M_{\odot}$)     & (K\,km\,s$^{-1}$)           &                                        & (K\,km\,s$^{-1}$)           &                       & ($^{\circ}$)      &                &                \\[-1.0mm]
\colhead{(1)} & \colhead{(2)}  & \colhead{(3)}   & \colhead{(4)}           & \colhead{(5)}           & \colhead{(6)}   & \colhead{(7)}        & \colhead{(8)}           & \colhead{(9)}           & \colhead{(10)}              & \colhead{(11)}                         & \colhead{(12)}              & \colhead{(13)}        & \colhead{(14)}    & \colhead{(15)} & \colhead{(16)}
}                                                                                                                                                                                                                                                                                                                                                
\startdata                                                                                                                                                                                                                                                                                                                                       
1	      & 23.46415       & 30.65744	 & 0.07			   & $-168$		     & 17.3 $\pm$ 4.4  & 3.7 $\pm$ 0.9	      &   3.8 $\pm$ 2.0	        & 24.2 $\pm$  8.6	  &  1.83 $\pm$ 0.08	        & 10.0	      				 & $<$ 0.36		       & $<$ 0.20	       & $ 24$	       	   & 5.0            & 149      \\
2	      & 23.46366       & 30.66494	 & 0.07			   & $-202$		     & 29.2 $\pm$ 4.2  & 4.6 $\pm$ 0.6        &  16.5 $\pm$ 3.3         & 64.4 $\pm$ 12.0	  &  6.50 $\pm$ 0.13	        & 18.2        				 & $<$ 0.35		       & $<$ 0.05	       & $-75$       	   & 2.1            &          \\
3	      & 23.46754       & 30.66369	 & 0.10			   & $-187$		     & 46.3 $\pm$ 10.2 & 2.5 $\pm$ 0.5        &  10.2 $\pm$ 4.5         & 30.8 $\pm$  8.8	  &  2.76 $\pm$ 0.12	        & 10.6        				 & $<$ 0.32		       & $<$ 0.11	       & $ 25$	       	   & 3.3            & 156      \\
4	      & 23.45834       & 30.65327	 & 0.11			   & $-162$		     & 19.0 $\pm$ 6.9  & 2.9 $\pm$ 0.9        &   2.5 $\pm$ 1.6         & 16.7 $\pm$  7.8	  &  1.46 $\pm$ 0.06	        & 12.4        				 & $<$ 0.25		       & $<$ 0.17	       & $-61$       	   & 1.4            &          \\
5	      & 23.46560       & 30.66952	 & 0.14			   & $-203$		     & 38.1 $\pm$ 3.9  & 2.9 $\pm$ 0.4        &  18.7 $\pm$ 3.8         & 32.4 $\pm$  6.1	  &  3.61 $\pm$ 0.09	        & 22.3        				 & $<$ 0.47		       & $<$ 0.13	       & $-62$       	   & 2.0            & 177      \\
6	      & 23.46754       & 30.65494	 & 0.16			   & $-168$		     & 48.2 $\pm$ 2.5  & 4.0 $\pm$ 0.2        & 125.4 $\pm$ 6.9         & 78.4 $\pm$  5.1	  & 20.94 $\pm$ 0.10	        & 95.8        				 & 2.64 $\pm$ 0.11	       & 0.13 $\pm$ 0.01       & $ 36$	       	   & 1.2            & 149      \\
7	      & 23.45495       & 30.65035	 & 0.17			   & $-163$		     & 12.6 $\pm$ 10.9 & 2.4 $\pm$ 1.0        &   1.1 $\pm$ 0.1         &  7.5 $\pm$  5.2	  &  1.20 $\pm$ 0.07	        & 7.5	      				 & $<$ 0.20		       & $<$ 0.17	       & $ 41$	       	   & 4.2            &          \\
8	      & 23.45301       & 30.65869	 & 0.17			   & $-165$		     & 35.7 $\pm$ 3.4  & 2.4 $\pm$ 0.2        &  18.7 $\pm$ 2.6         & 21.8 $\pm$  3.1	  &  7.37 $\pm$ 0.07	        & 41.3        				 & 0.52 $\pm$ 0.11	       & 0.07 $\pm$ 0.01       & $-86$        	   & 2.3            & 151      \\
9	      & 23.45543       & 30.66535	 & 0.18			   & $-177$		     & 39.4 $\pm$ 6.7  & 4.6 $\pm$ 0.7        &   9.2 $\pm$ 0.5         & 88.4 $\pm$ 18.4	  &  2.62 $\pm$ 0.10	        & 14.9        				 & $<$ 0.39		       & $<$ 0.15	       & $-27$         	   & 1.3            & 160      \\
10	      & 23.45494       & 30.66452	 & 0.18			   & $-187$		     & 25.9 $\pm$ 6.0  & 2.0 $\pm$ 0.5        &   3.7 $\pm$ 2.2         & 10.5 $\pm$  4.0	  &  3.15 $\pm$ 0.09	        & 13.5        				 & $<$ 0.23		       & $<$ 0.07	       & $-64$       	   & 1.9            & 160      \\
$\cdots$      & $\cdots$       & $\cdots$        & $\cdots$                & $\cdots$                & $\cdots$        & $\cdots$             & $\cdots$                & $\cdots$                & $\cdots$                    & $\cdots$    				 & $\cdots$                    & $\cdots$              & $\cdots$          & $\cdots$       & $\cdots$ \\
$\cdots$      & $\cdots$       & $\cdots$        & $\cdots$                & $\cdots$                & $\cdots$        & $\cdots$             & $\cdots$                & $\cdots$                & $\cdots$                    & $\cdots$    				 & $\cdots$                    & $\cdots$              & $\cdots$          & $\cdots$       & $\cdots$ \\
839	      & 23.65535       & 30.58110	 & 4.69			   & $-174$		     & 29.5 $\pm$ 8.8  & 2.5 $\pm$ 0.5	      &   7.6 $\pm$ 3.3	        & 19.4 $\pm$  5.5	  &  3.00 $\pm$ 0.12	        & 12.0	      				 & $<$ 0.37		       & $<$ 0.12	       & $ 43$	           & 1.2            & 326      \\
840	      & 23.65583       & 30.57610	 & 4.74			   & $-176$		     & 32.2 $\pm$ 6.8  & 2.2 $\pm$ 0.3        &   8.7 $\pm$ 3.5         & 16.7 $\pm$  4.0 	  &  3.47 $\pm$ 0.09	        & 19.3        				 & $<$ 0.33		       & $<$ 0.10	       & $ 69$	       	   & 1.5            & 326      \\
841	      & 23.65965       & 30.54693	 & 5.05			   & $-154$		     & 13.3 $\pm$ 9.3  & 1.8 $\pm$ 0.8        &   1.7 $\pm$ 1.1         &  4.5 $\pm$  3.2 	  &  1.38 $\pm$ 0.10	        & 7.1	      				 & $<$ 0.18		       & $<$ 0.13	       & $-15$       	   & 5.7            &          \\
842	      & 23.66347       & 30.51901	 & 5.36			   & $-152$		     & 38.6 $\pm$ 6.4  & 2.7 $\pm$ 0.4        &  13.2 $\pm$ 4.8         & 30.0 $\pm$  6.7 	  &  2.49 $\pm$ 0.09	        & 16.5        				 & $<$ 0.34		       & $<$ 0.14	       & $-56$       	   & 3.9            & 252      \\
843	      & 23.67514       & 30.54941	 & 5.36			   & $-165$		     & 17.2 $\pm$ 10.2 & 1.8 $\pm$ 0.7        &   1.7 $\pm$ 1.8         &  5.5 $\pm$  3.7 	  &  1.08 $\pm$ 0.07	        & 8.0	      				 & $<$ 0.35		       & $<$ 0.33	       & $ 62$	       	   & 2.0            &          \\
844	      & 23.67173       & 30.53649	 & 5.39			   & $-161$		     & 16.7 $\pm$ 8.7  & 1.8 $\pm$ 0.6        &   1.8 $\pm$ 0.9         &  5.6 $\pm$  3.2 	  &  1.25 $\pm$ 0.08	        & 9.3	      				 & $<$ 0.27		       & $<$ 0.22	       & $ 27$	       	   & 1.5            &          \\
845	      & 23.67512       & 30.54024	 & 5.43			   & $-158$		     & 24.9 $\pm$ 4.2  & 2.1 $\pm$ 0.3        &  17.2 $\pm$ 3.6         & 11.3 $\pm$  2.6 	  &  6.57 $\pm$ 0.12	        & 30.8        				 & 0.84 $\pm$ 0.17	       & 0.13 $\pm$ 0.03       & $  3$	       	   & 1.3            & 291      \\
846	      & 23.67899       & 30.54273	 & 5.50			   & $-164$		     & 35.6 $\pm$ 9.6  & 2.0 $\pm$ 0.4        &   6.4 $\pm$ 1.7         & 14.1 $\pm$  4.5 	  &  2.08 $\pm$ 0.08	        & 15.5        				 & $<$ 0.46		       & $<$ 0.22	       & $ 73$	       	   & 3.2            & 291      \\
847	      & 23.67461       & 30.52941	 & 5.51			   & $-156$		     & 42.1 $\pm$ 7.6  & 2.1 $\pm$ 0.4        &   8.1 $\pm$ 1.5         & 18.7 $\pm$  5.3 	  &  1.74 $\pm$ 0.09	        & 11.9        				 & $<$ 0.26		       & $<$ 0.15	       & $-66$        	   & 2.0            & 290      \\
848	      & 23.67362       & 30.51608	 & 5.59			   & $-155$		     & 36.6 $\pm$ 14.4 & 2.2 $\pm$ 0.3        &   9.2 $\pm$ 0.9         & 18.2 $\pm$  5.2 	  &  3.10 $\pm$ 0.15	        & 11.2        				 & $<$ 0.37		       & $<$ 0.12	       & $-75$       	   & 1.2            & 253      \\
\enddata                                                                                                                                                                                                                                                                                                 
\tablecomments{
(1) ID number of the cloud.
(2) -- (3) $^{12}$CO($J=2-1$) peak position of the cloud in equatorial coordinates (J2000) in degree.
(4) Galactocentric radius of the cloud from the optical center of M33 (1$^{\rm h}$33$^{\rm m}$50$^{\rm s}$.9, 30$^{\circ}$39\arcmin37\arcsec) in units of kiloparsec.
(5) Radial velocity in the Local Standard of Rest in units of km\,s$^{-1}$.
(6) Deconvolved radius of the cloud including uncertainty in units of parsec.
(7) Deconvolved velocity dispersion including uncertainty in units of km\,s$^{-1}$.
(8) Luminosity mass based on $^{12}$CO($J=2-1$) flux including uncertainty in units of $10^4\,M_{\odot}$.
(9) Mass of the cloud inferred from the virial theorem including uncertainty in units of $10^4\,M_{\odot}$.
(10) $^{12}$CO($J=2-1$) intensity at its peak position of the cloud including uncertainty in units of K\,km\,s$^{-1}$.
(11) S/N ratio of $^{12}$CO($J=2-1$) brightness temperature at its peak position.
(12) $^{13}$CO($J=2-1$) intensity at $^{12}$CO($J=2-1$) peak position of the cloud including uncertainty in units of K\,km\,s$^{-1}$, or its 3\,$\sigma$ upper limit.
(13) $^{13}$CO($J=2-1$)/$^{12}$CO($J=2-1$) intensity ratio at $^{12}$CO($J=2-1$) peak position of the cloud including uncertainty, or its 3\,$\sigma$ upper limit.
(14) Position angle of the major axis of the cloud, which is measured counterclockwise from north to east, in units of degree.
(15) Ratio between the major and minor axes after the deconvolution by the observing beam (7\farcs5).
(16) GMC counterpart identified with the IRAM 30\,m telescope \citep{corbelli2017}.
}
\end{deluxetable*}
\end{longrotatetable}

\subsection{Basic Properties of Cataloged Molecular Clouds} \label{sec:prop}

\begin{figure*}[ht!]
\epsscale{0.5}
\plotone{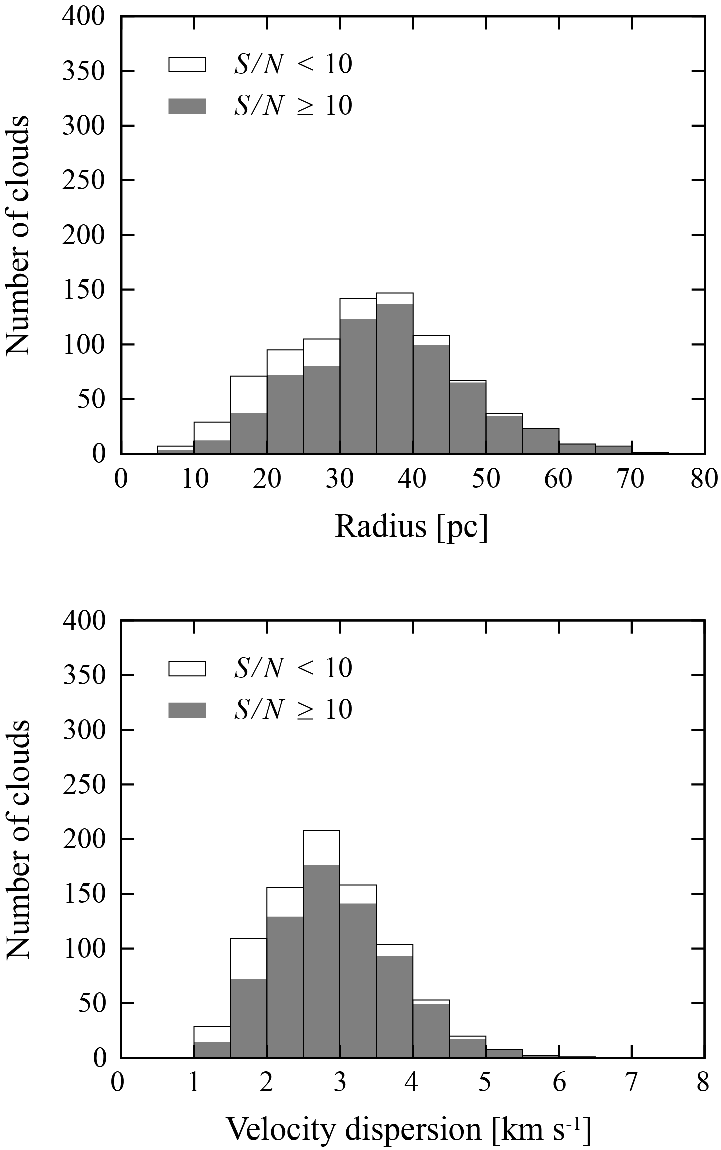}
\caption{Frequency distributions of the radius ($R$; top) and the velocity dispersion ($\sigma_v$; bottom) for the cataloged molecular clouds in M33.
A shaded column indicates high-S/N clouds and a white column corresponds to low-S/N clouds.
}\label{fig:histo-1}
\end{figure*}

\begin{figure*}[ht!]
\epsscale{0.5}
\plotone{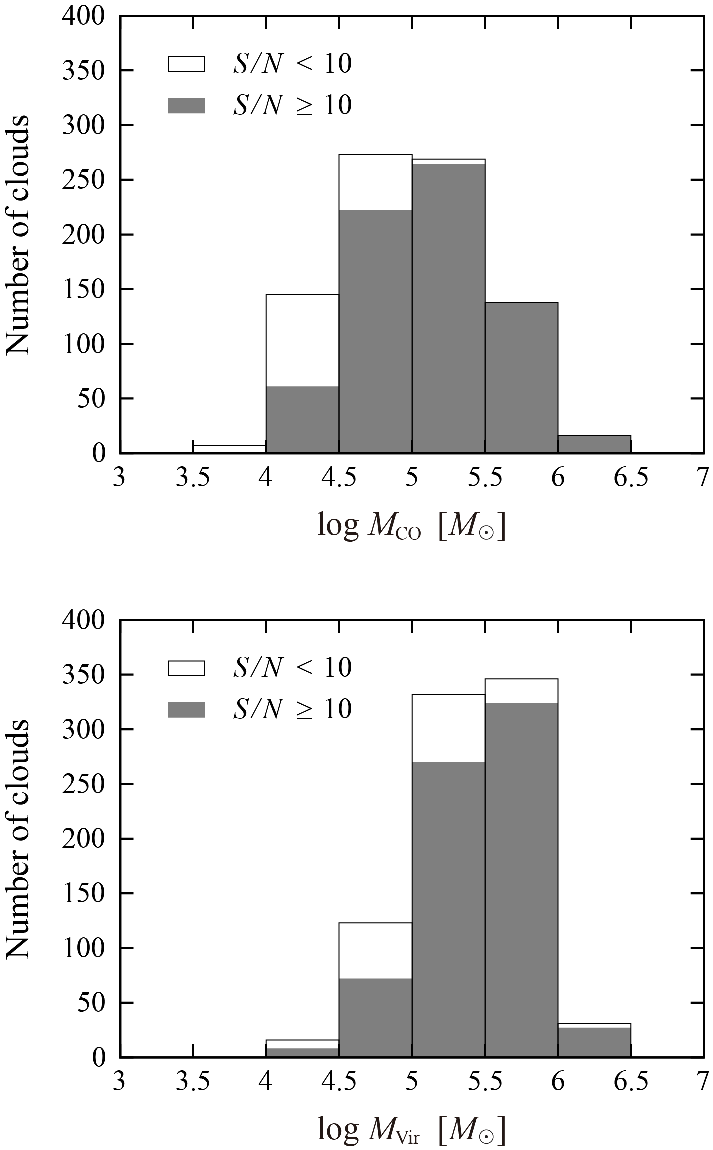}
\caption{Frequency distributions of the $^{12}$CO luminosity-based mass ($M_{\rm CO}$; top) and the virial mass ($M_{\rm Vir}$; bottom) for the cataloged molecular clouds in M33.
A shaded column indicates high-S/N clouds and a white column corresponds to low-S/N clouds.
}\label{fig:histo-2}
\end{figure*}

Figure~\ref{fig:histo-1} shows the frequency distributions of the radius $R$ and the velocity dispersion $\sigma_v$ for the cataloged molecular clouds in M33.
$R$ ranges from 6.8 to 72\,pc, and $\sigma_v$ ranges from 1.0 to 6.1\,km\,s$^{-1}$. Their medians are 34\,pc and 2.8\,km\,s$^{-1}$, respectively.
Note that, as pointed out by \cite{hughes2013}, such distributions of the cloud radius and the velocity dispersion depend on both the spatial and the velocity resolutions of the input data cube
because ISM in galaxies generally has a hierarchical structure from parsec to kiloparsec scales.
We also examined the frequency distributions of $M_{\rm CO}$ and $M_{\rm Vir}$ as shown in Figure~\ref{fig:histo-2}.
$M_{\rm CO}$ ranges from $6.7 \times 10^3$ to $2.6 \times 10^6\,M_{\odot}$, and $M_{\rm Vir}$ ranges from $1.1 \times 10^4$ to $1.9 \times 10^6\,M_{\odot}$.
Their medians are $9.9 \times 10^4\,M_{\odot}$ and $2.8 \times 10^5\,M_{\odot}$, respectively.
Both for $M_{\rm CO}$ and $M_{\rm Vir}$, the dynamic range of mass is more than two orders of magnitude, which is wider than earlier M33 studies \citep[e.g.,][]{rosolowsky2007, gratier2012, corbelli2017}.
The low-S/N ($< 10$) clouds typically show smaller radii and smaller velocity dispersions compared to the high-S/N clouds.
However, some low-S/N clouds have large virial masses ($\geq 10^5 M_{\odot}$) although their CO luminosity masses are almost small ($< 10^5 M_{\odot}$).
We discuss the origin of the discrepancy between $M_{\rm CO}$ and $M_{\rm Vir}$ for the low-S/N clouds in subsection~\ref{sec:massrelation}.

\section{Scaling Relations} \label{sec:scaling}

Starting with the pioneering work by \cite{larson1981}, a lot of earlier studies suggest that basic properties of molecular clouds are quantitatively related to each other via some kind of scaling relations,
which are often referred to as ``Larson's laws''.
In this section, we examine such scaling relations based on our molecular-cloud catalog.

\subsection{Size--Line-width Relation} \label{sec:sizewidth}

\begin{figure*}[ht!]
\epsscale{0.5}
\plotone{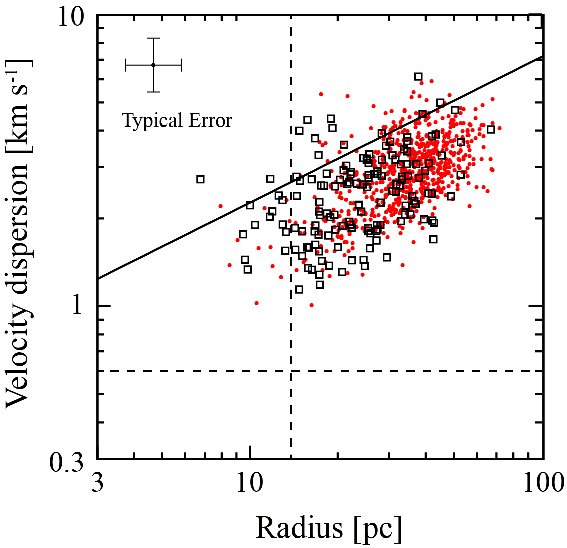}
\caption{Radius-velocity dispersion ($R-\sigma_v$) relation for the molecular clouds in M33.
Red dots indicate high-S/N clouds and open squares indicate low-S/N clouds.
The solid line indicates the relation for the Galactic clouds \citep[$\sigma_v$ = 0.72$R^{0.5}$:][]{solomon1987}.
The dashed vertical and horizontal lines indicate the resolution limits obtained from the spatial and velocity resolutions of the ACA+IRAM $^{12}$CO($J=2-1$) data.
}\label{fig:sizewidth}
\end{figure*}

The first Larson's law relates the cloud radius $R$ in parsecs to the velocity dispersion $\sigma_v$ in km\,s$^{-1}$,
which is expressed as $\sigma_v = 0.72 R^{0.5}$ for the MW \citep{solomon1987}.
This relation is considered to reflect the turbulent condition inside the molecular clouds.
Figure~\ref{fig:sizewidth} shows the relation between $R$ and $\sigma_v$ for 848 molecular clouds in M33, distinguishing the low-S/N and high-S/N clouds.
The overall distributions in the radius-velocity dispersion plane are similar between the two cloud types;
many clouds show smaller velocity dispersion than the Galactic $R-\sigma_v$ relation at a given radius regardless of the $^{12}$CO S/N ratio.
This trend can be evaluated more quantitatively by deriving the coefficient of the $R-\sigma_v$ relation, $\sigma_v R^{0.5}$.
The average $\sigma_v R^{0.5}$ for 848 clouds is $0.48 \pm 0.13$, which is significantly smaller than the Galactic $\sigma_v R^{0.5}$, 0.72.

To explain the origin of such a smaller velocity dispersion, we examine the $R-\sigma_v$ relations based on the two earlier GMC catalogs in M33 \citep{rosolowsky2007, corbelli2017}.
We found that the averaged $\sigma_v R^{0.5}$ are $0.54 \pm 0.11$ and $0.50 \pm 0.17$ for the \cite{rosolowsky2007} catalog and the \cite{corbelli2017} catalog, respectively,
although the linewidth of small ($<$ 20\,pc) clouds in the \cite{corbelli2017} catalog is comparable to the MW as pointed out by \cite{braine2018}.
This suggests that the velocity dispersion of the GMC in M33 is intrinsically smaller than the Galactic GMCs.

Here, we consider the physical mechanism to change the velocity dispersion in molecular clouds.
Earlier studies reported that the velocity dispersion at a given cloud radius is higher in the Galactic Center \citep{oka2001} and 30~Doradus in the LMC \citep{wong2017, wong2019},
which are associated with active star-forming regions, compared to the Galactic $R-\sigma_v$ relation \citep{solomon1987}.
In contrast to this, the quiescent cloud PGCC G282.92-32.40 \citep{planck2016}, which is referred to as the ``Planck Cold Cloud (PCC)'', in the LMC shows a smaller velocity dispersion than the Galactic $R-\sigma_v$ relation \citep{wong2017, wong2019}.
From these observational facts, \cite{wong2019} suggested that local energy injection by star formation feedback plays an important role in the turbulence of molecular clouds.
In other words, it is unlikely that a molecular cloud without active star-forming regions increases its velocity dispersion.
Considering that the $R-\sigma_v$ relation obtained in M33 is similar to that in the PCC,
it is suggested that many molecular clouds in M33 are not associated with active star formation like the PCC.
However, this contradicts the fact that more than 70\% of clouds in M33 are associated with star-forming regions \citep[e.g.,][]{gratier2012, konishi2023}.
Indeed, Figure~\ref{fig:cloudsmap} shows that many high-mass clouds are associated with 8\,$\mu$m-bright sources (see also Figure~\ref{fig:cdf_8um}).
Alternatively, \cite{bolatto2008} examined the variation in $\sigma_v R^{0.5}$ for 12 external galaxies
and found a trend that extragalactic GMCs falling under the Galactic $R-\sigma_v$ relation have lower surface densities ($\Sigma_{\rm GMC}$) than corresponding clouds in the MW.
GMCs in SMC and NGC\,4605 show $\Sigma_{\rm GMC} \sim 45$ and $\sigma_v R^{0.5} = 0.37$.
Both values are lower than GMCs in other 10 external galaxies, while similar to those in M33 ($\Sigma_{\rm GMC} = 30 - 40$ and $\sigma_v R^{0.5} = 0.4 - 0.5$).
This suggests that low-surface density molecular clouds can be maintained even by the small turbulence (i.e., small velocity dispersion), which results in the observed $R-\sigma_v$ relation in M33 \citep[see also][]{ohno2023}.

\subsection{CO Luminosity Mass--Virial Mass Relation} \label{sec:massrelation}

\begin{figure*}[ht!]
\epsscale{0.5}
\plotone{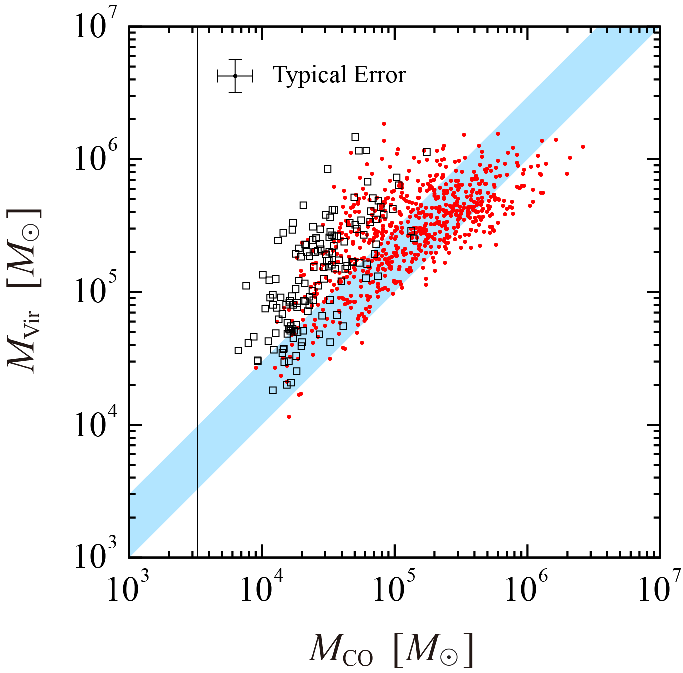}
\caption{Comparison between the virial mass ($M_{\rm Vir}$) and the $^{12}$CO luminosity-based mass ($M_{\rm CO}$) for the molecular clouds in M33.
Red dots indicate high-S/N clouds and open squares indicate low-S/N clouds.
The filled area in blue indicates $1 \leq M_{\rm Vir}/M_{\rm CO} \leq 3$, in which clouds are in virial equilibrium.
The vertical line indicates the detection limit derived from the emission masks and \texttt{PYCPROPS} parameters.
}\label{fig:mcomvir}
\end{figure*}

Figure~\ref{fig:mcomvir} shows a relationship between $M_{\rm CO}$ and $M_{\rm Vir}$ for 848 molecular clouds in M33.
Both masses seem to be well correlated, whereas $M_{\rm Vir}$ is generally larger than $M_{\rm CO}$. A similar trend is also reported by \cite{corbelli2017}.
In particular, the low-S/N clouds show larger $M_{\rm Vir}$ at a given $M_{\rm CO}$; the median of the virial parameter $\alpha$, which is defined as $M_{\rm Vir}/M_{\rm CO}$, is 5.3 for the low-S/N clouds.
Since clouds in virial equilibrium show $\alpha \approx$ 1\,--\,3 \citep{miville2017}, most of the low-S/N clouds are gravitationally unbound ($\alpha > 3$).
On the other hand, the median of $\alpha$ for the high-S/N clouds is 2.0, and 68\% of the high-S/N clouds are gravitationally bound ($\alpha \leq 3$).
Considering that the $R-\sigma_v$ relation is not different between the high-S/N clouds and low-S/N clouds, the CO intensity emitted from the low-S/N cloud may be simply weak.
This is consistent with the low-surface density molecular clouds in M33 (see subsection~\ref{sec:sizewidth}).
In Figure~\ref{fig:mcomvir}, we can see most of the clouds are virialized ($\alpha \leq 3$) at the high-mass end.
We discuss the physical meaning of this trend in Section~\ref{sec:sf}.

\section{Comparison with Earlier GMC catalog} \label{sec:comparison}

Based on the ACA+IRAM $^{12}$CO($J=2-1$) data of M33, we cataloged 848 molecular clouds.
In this section, we compare the ACA+IRAM cloud catalog with the earlier GMC catalog generated by \cite{corbelli2017}.
To give a fair comparison between the two catalogs, we extracted 362 GMCs from the \cite{corbelli2017} catalog which are located within the ACA FOV.
Hereafter, we refer to the GMCs in the \cite{corbelli2017} catalog as ``IRAM GMCs''.
This comparison between the two catalogs provides new insights into the hierarchical structure of molecular gas.

\subsection{Mass Function} \label{sec:massfunc}

Firstly, we investigate the cloud mass distributions for the two catalogs.
The cumulative mass distribution function can be expressed by truncated power-law functions as follows:

\begin{eqnarray}
N(M' > M) = N_0 \left[ \left( \frac{M}{M_0} \right)^{\gamma +1} -1 \right],
\end{eqnarray}
where $M_0$ is the maximum mass in the distribution, $\gamma$ indicates how the cloud mass is distributed, and $N_0$ is the number of clouds more massive than $2^{1/(\gamma+1)}M_0$ \citep[e.g.,][]{rosolowsky2005}.
To determine the fitting range of the cloud mass distributions, we estimated the completeness limit of molecular clouds by reference to \cite{engargiola2003}.
They reported that the lowest mass molecular cloud in their GMC survey is of order $2 \times 10^4\,M_{\odot}$ and also estimated the completeness limit of $1.5 \times 10^5\,M_{\odot}$, which is about seven times larger than the lowest mass.
If we apply such a linear scaling between the two masses to our molecular cloud catalog, the completeness limit is estimated to be $5 \times 10^4\,M_{\odot}$ because the lowest mass cloud is $7 \times 10^3\,M_{\odot}$ (see subsection~\ref{sec:masslumi}).
Note that we recalculated the GMC masses in the \cite{corbelli2017} catalog by assuming $R_{21} = 0.6$ and adopted $8.4 \times 10^4\,M_{\odot}$ as the completeness limit\footnote{\cite{corbelli2017} reported that the completeness limit is $6.3 \times 10^4\,M_{\odot}$ in their GMC catalog. The assumption of $R_{21} = 0.6$ in this study yields the corrected completeness limit of $6.3 \times 10^4 \times (0.8/0.6) = 8.4 \times 10^4\,M_{\odot}$.}.

\begin{figure*}[ht!]
\epsscale{1.15}
\plotone{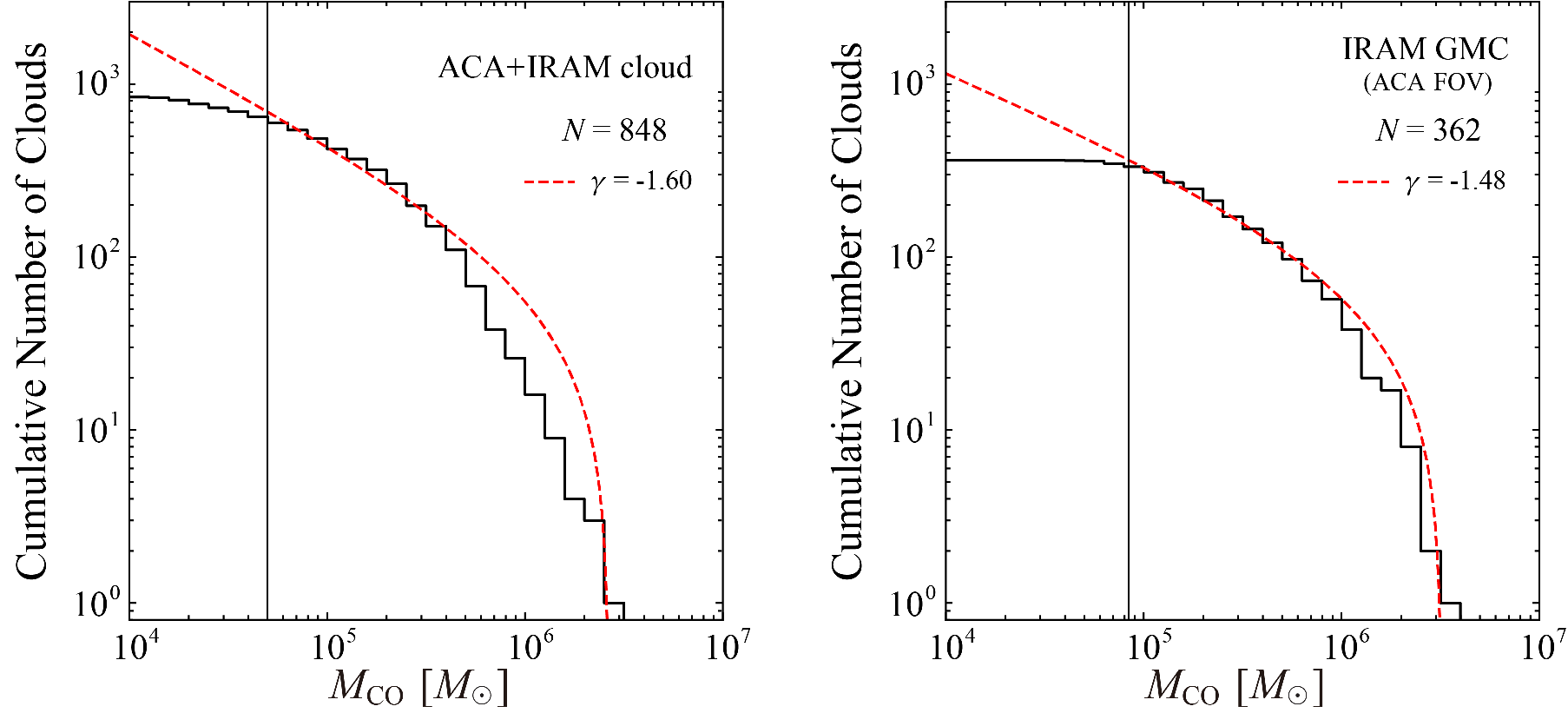}
\caption{Cumulative cloud mass functions from the ACA+IRAM catalog (left) and the \cite{corbelli2017} catalog (right).
The red dashed lines indicate the fitting results by the truncated power-law functions.
The vertical lines show the completeness limit in each plot.
}\label{fig:massfunc_all}
\end{figure*}

\begin{figure*}[ht!]
\epsscale{0.8}
\plotone{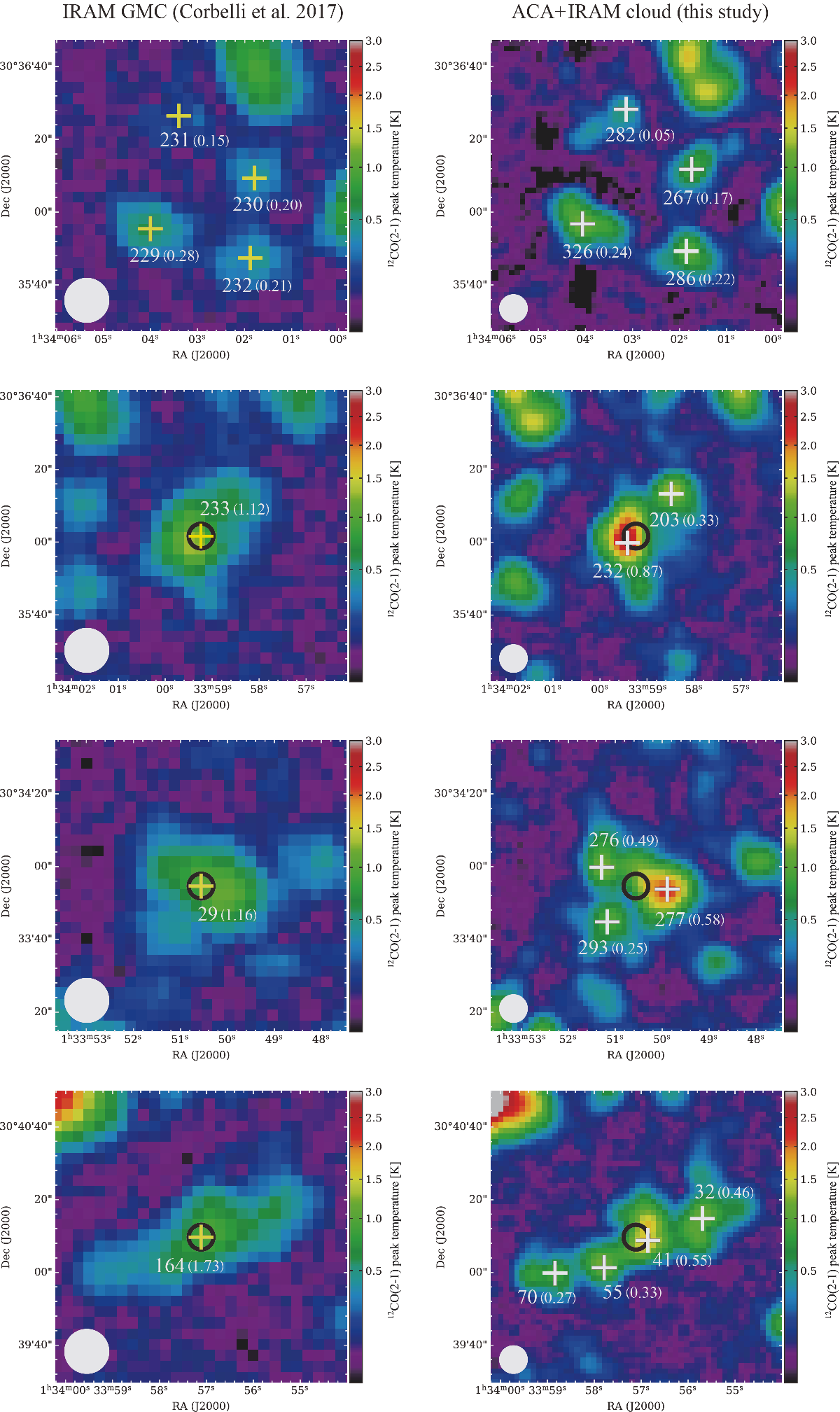}
\caption{Comparison of the cloud identification between IRAM GMCs (left) and ACA+IRAM clouds (right) in $^{12}$CO($J=2-1$) peak temperature maps.
Yellow crosses and black circles indicate $^{12}$CO($J=2-1$) peak positions in each IRAM GMCs.
White crosses indicate $^{12}$CO($J=2-1$) peak positions in each ACA+IRAM cloud.
Numbers in the left column correspond to GMC IDs in the IRAM GMC catalog \citep{corbelli2017}, and those in the right column are cloud IDs in the ACA+IRAM cloud catalog (Table~\ref{tab:clouds}).
The value in parentheses after the ID indicates $M_{\rm CO}$ in units of $10^6\,M_{\odot}$ for each cloud.
The beam size is shown in the lower left corner of each map.
}\label{fig:resolved}
\end{figure*}

Figure~\ref{fig:massfunc_all} shows the cumulative cloud mass functions fitted by the truncated power-law functions.
Note that we treated $N_0$ as a free parameter (i.e., independent of $\gamma$) in this study to achieve the best fitting.
We obtained $\gamma = -1.60$ for the ACA+IRAM cloud catalog and $\gamma = -1.48$ for the \cite{corbelli2017} catalog, respectively.
Considering that \cite{braine2018} obtained $\gamma = -1.65$ for all 566 GMCs in the \cite{corbelli2017} catalog, the cloud population of the ACA+IRAM cloud catalog is similar to that of the \cite{corbelli2017} catalog.
However, the truncated power-law fitting for the ACA+IRAM cloud catalog deviates from the mass spectrum at the high-mass side (especially from $5 \times 10^5\,M_{\odot}$ to $2 \times 10^6\,M_{\odot}$);
the number of clouds in this mass range is significantly less than the expectation by the truncated power-law function and also less than the \cite{corbelli2017} catalog.
Such a decrease in the high-mass clouds in the ACA+IRAM cloud catalog is presumably due to the difference in the spatial resolutions of CO data;
some large IRAM GMCs identified with the 49\,pc beam can be resolved into multiple cloud components in the ACA+IRAM CO data at 30\,pc resolution.
This yields a decrease in the number of GMCs at the high-mass side.
Indeed, we examined a one-on-one comparison between ACA+IRAM clouds and IRAM GMCs and found that 170 IRAM GMCs are resolved into 2 or more ACA+IRAM clouds.
Figure~\ref{fig:resolved} shows the comparison of the cloud identification between the two CO data.
A small IRAM GMC is identified as a single molecular cloud even in the ACA+IRAM CO data, while a large IRAM GMC are resolved into multiple ACA+IRAM clouds.

\subsection{Origin of Velocity Dispersion in IRAM GMCs} \label{sec:sigmav}

As described above, a large IRAM GMC (typically its $M_{\rm CO}$ is larger than $3 \times 10^5 M_{\odot}$) can be treated as an association of multiple ACA+IRAM clouds.
Investigating such a correspondence is beneficial for the comparison between properties of individual molecular clouds and the average properties of their association.
In particular, we focus on the origin of the observed velocity dispersion (linewidth) of a large IRAM GMC, which is likely composed of two factors;
(1) the line-of-sight relative velocity between internal ACA+IRAM clouds and (2) velocity dispersions of individual ACA+IRAM clouds.
Here we examine which factors mainly contribute to the overall velocity dispersion for 77 IRAM GMCs, which are resolved into 3 or more ACA+IRAM clouds.

To quantify the line-of-sight velocity difference between multiple ACA+IRAM clouds, 
we firstly defined the weighted center of line-of-sight velocities between the clouds as follows:

\begin{eqnarray}
v_g = \frac{\sum\limits_{i=1}^n I_i\,v_i}{\sum\limits_{i=1}^n I_i},
\end{eqnarray}
where $I_i$ and $v_i$ are the $^{12}$CO($J=2-1$) intensity at the CO peak position (10th column in Table~\ref{tab:clouds}) and the line-of-sight velocity ($V_{\rm LSR}$; 5th column in Table~\ref{tab:clouds}) of $i$th ACA+IRAM cloud, respectively.
$n$ is the number of ACA+IRAM clouds included in a large IRAM GMC.
Using this $v_g$, we calculate the representative velocity difference between internal ACA+IRAM clouds as follows:

\begin{eqnarray}
v_{\rm diff} = \sqrt{\frac{\sum\limits_{i=1}^n I_i (v_i - v_g)^2}{\sum\limits_{i=1}^n I_i}}.
\end{eqnarray}
In addition, we calculate the weighted mean of velocity dispersions of individual ACA+IRAM clouds as follows:

\begin{eqnarray}
\sigma_{v, {\rm mean}} = \frac{\sum\limits_{i=1}^n I_i\,\sigma_{v, i}}{\sum\limits_{i=1}^n I_i},
\end{eqnarray}
where $\sigma_{v, i}$ is the velocity dispersion of $i$th ACA+IRAM cloud.

\begin{figure*}[ht!]
\epsscale{1}
\plotone{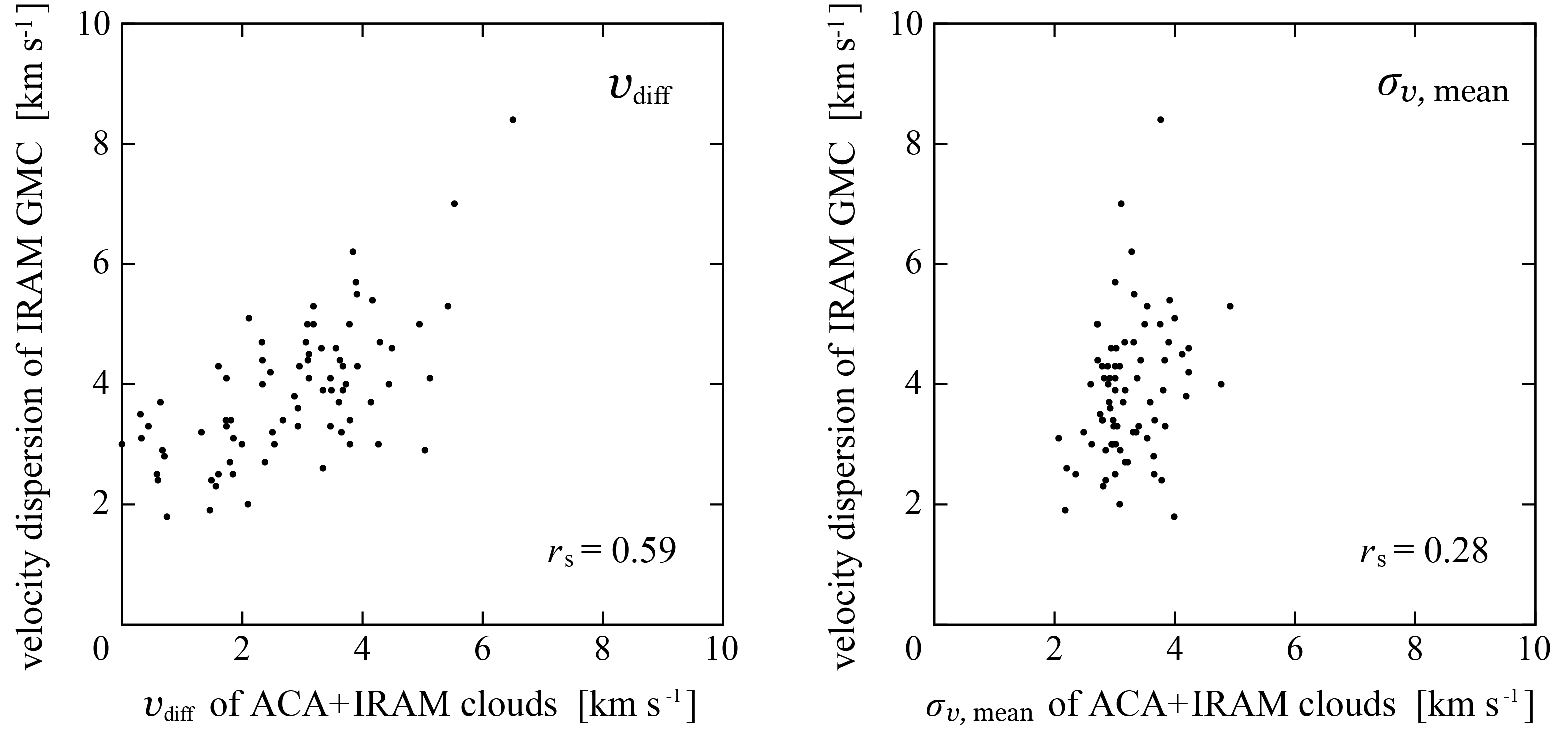}
\caption{Velocity dispersion of the IRAM GMC as a function of the representative line-of-sight velocity difference between internal ACA+IRAM clouds $v_{\rm diff}$ (left) and that of the weighted mean of velocity dispersions of individual ACA+IRAM clouds $\sigma_{v, {\rm mean}}$ (right).
The Spearman's rank correlation coefficient $r_{\rm s}$ is shown in the bottom-right corner of each plot.
}\label{fig:dispersion}
\end{figure*}

Figure~\ref{fig:dispersion} shows the velocity dispersion of the IRAM GMC as a function of $v_{\rm diff}$ and that of $\sigma_{v, {\rm mean}}$.
A clear correlation between $v_{\rm diff}$ and the velocity dispersion of the IRAM GMC, with the Spearman's rank correlation coefficient $r_{\rm s}$ of 0.59, can be seen,
while $\sigma_{v, {\rm mean}}$ seems nearly constant (2\,--\,4\,km\,s$^{-1}$) and its correlation with the velocity dispersion of the IRAM GMC is weak ($r_{\rm s}$ = 0.28).
This suggests that the velocity dispersion of a large cloud is mainly dominated by the line-of-sight velocity difference between small clouds inside the GMC in the case of $v_{\rm diff} >$ 2\,km\,s$^{-1}$,
while the velocity dispersion of individual internal clouds determines the overall velocity dispersion of the GMC if $v_{\rm diff}$ is less than 2\,km\,s$^{-1}$.

\section{Properties of Molecular Clouds and High-mass Star Formation} \label{sec:sf}

\begin{figure*}[ht!]
\epsscale{0.5}
\plotone{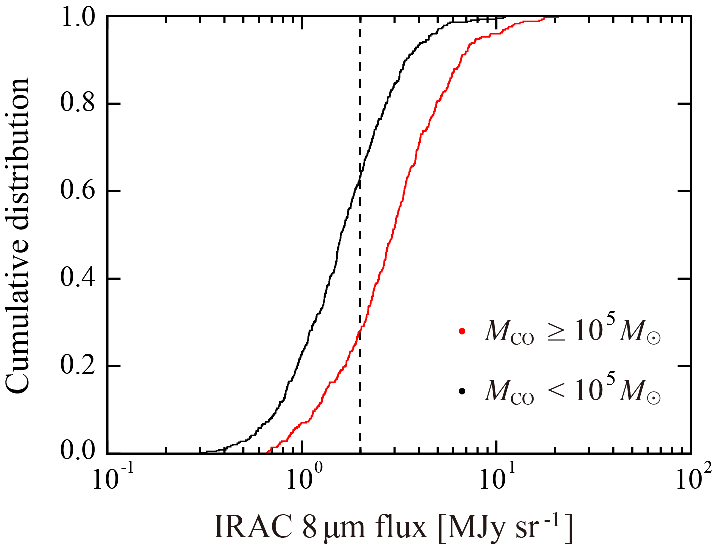}
\caption{Normalized CDF of mean $8\,\mu$m flux in each molecular cloud.
The red and black lines indicate the CDF for 423 high-mass clouds ($M_{\rm CO} \geq 10^5\,M_{\odot}$) and 425 low-mass clouds ($M_{\rm CO} < 10^5\,M_{\odot}$), respectively.
The vertical dashed line indicates the $8\,\mu$m flux of 2\,MJy\,sr$^{-1}$.
}\label{fig:cdf_8um}
\end{figure*}

As shown in Figure~\ref{fig:cloudsmap} (and also described in subsection~\ref{sec:masslumi}), 
many high-mass clouds are associated with the strong 8\,$\mu$m emission in the spiral arm region,
while low-mass clouds tend to be apart from such $8\,\mu$m-bright sources and to exist in the inter-arm region.
Here, we quantitatively evaluate the relationship between the molecular clouds and the 8\,$\mu$m-bright sources.
To do this, we regridded the IRAC 8\,$\mu$m map \citep{dale2009} to match the ACA+IRAM $^{12}$CO($J=2-1$) map and obtained mean $8\,\mu$m flux by averaging the pixel values included within each molecular cloud.
Then, we constructed the cumulative distribution functions (CDFs) both for 423 high-mass clouds and 425 low-mass clouds.
Figure~\ref{fig:cdf_8um} clearly shows that the $8\,\mu$m-bright sources are closely associated with high-mass clouds rather than low-mass clouds;
the strong ($> 2$\,MJy\,sr$^{-1}$) $8\,\mu$m emission is found in 72\% of high-mass clouds, but only in 36\% of low-mass clouds, respectively.
Note that this trend does not change even if we exclude the diffuse components of 8\,$\mu$m emission (see appendix).
Our result indicates that high-mass star formation tends to be associated with high-mass clouds rather than low-mass clouds.
Such a trend is consistent with the extensive study by \cite{corbelli2017}; they identified mid-infrared (MIR) emission with GMCs and found that
a GMC with bright MIR sources tends to have a large CO luminosity mass.

\begin{figure*}[ht!]
\epsscale{0.5}
\plotone{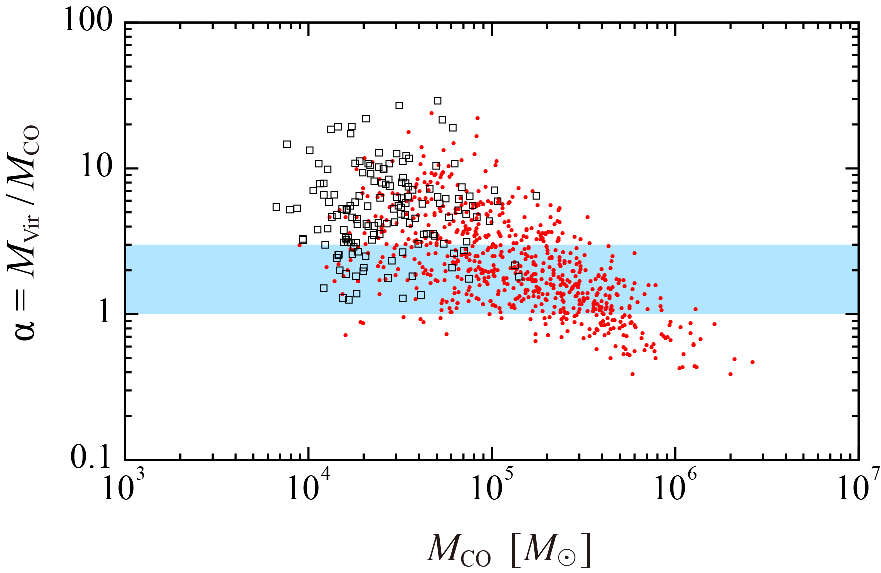}
\caption{Virial parameter $\alpha$ as a function of the $^{12}$CO luminosity-based mass ($M_{\rm CO}$).
The filled area in blue indicates $1 \leq \alpha \leq 3$, in which clouds are in virial equilibrium.
Red dots indicate high-S/N clouds and open squares indicate low-S/N clouds.
}\label{fig:alpha}
\end{figure*}

Since high-mass star formation generally starts from the gravitational instability of molecular gas,
the virial parameter $\alpha$, which expresses the degree of gravitational binding, is useful to examine star formation in molecular clouds.
Figure~\ref{fig:alpha} shows $\alpha$ as a function of $M_{\rm CO}$ for each ACA+IRAM cloud in M33.
$\alpha$ generally decreases (i.e., becomes more unstable against gravitational collapse) with the increase in $M_{\rm CO}$.
A similar trend is also observed in the MW \citep[e.g.,][]{miville2017} and external galaxies \citep[IC\,342;][]{hirota2011}.
In M33, high-mass clouds whose mass is larger than $10^5$\,$M_{\odot}$ seem to be almost virialized;
in other words, the self-gravitation is predominant rather than the internal turbulence of the cloud.
This indicates that the high-mass star formation likely onsets within such high-mass clouds by the gravitational instability,
which is well consistent with the observed feature; many high-mass clouds are associated with $8\,\mu$m-bright sources (Figures~\ref{fig:cloudsmap} and \ref{fig:cdf_8um}).
In addition, a large $\alpha$ for the low-S/N clouds with a median of 5.3 can be explained;
the low-S/N clouds largely correspond to low-mass clouds, which are gravitationally unbound and not associated with star-forming regions.

Finally, we briefly discuss the evolution of molecular clouds.
In M33, many high-mass clouds exist in the spiral arm region, while the inter-arm region is dominated by low-mass clouds.
Since the $8\,\mu$m-bright sources are loosely along spiral arms in M33, the stellar potential may play a vital role in the accumulation (and the resultant mass growth) of molecular clouds.
The evolution of molecular clouds crossing the spiral arm and the high-mass star formation within them are often discussed for grand-design spiral galaxies such as M51 \citep{egusa2011} and IC\,342 \citep{hirota2011}
based on the interferometric CO($J=1-0$) observations at a spatial resolution of a few $\times$ 10\,pc.
In M51, \cite{egusa2011} suggested that smaller molecular clouds collide to form smooth giant molecular associations (GMAs) at spiral arm regions and then star formation is triggered in the GMA cores.
\cite{hirota2011} divided the GMCs in the spiral arm of IC\,342 into two categories according to whether they are associated with star formation activity or not,
and reported that the GMCs with H\,{\sc ii} regions are typically more virialized and massive compared to the GMCs without H\,{\sc ii} regions.
These results are consistent with the picture of molecular clouds and the high-mass star formation in M33 although it is a flocculent galaxy whose spiral arm structures are relatively weak.
In a forthcoming paper, we will report a detailed study on the evolutionary stage of GMCs based on the comparison with H\,{\sc ii} regions \citep{konishi2023}.
Although the GMC evolution in M33 was investigated in earlier studies \citep[e.g.,][]{miura2012, corbelli2017}, the new ACA CO($J=2-1$) data enable us
to study the dense-gas formation based on $^{13}$CO($J=2-1$) emission as well as the evolution of basic properties of clouds (e.g., size, linewidth, mass, and virial parameters).

\section{Summary} \label{res:sum}

We have performed ALMA-ACA 7\,m-array observations in $^{12}$CO($J=2-1$), $^{13}$CO($J=2-1$), and C$^{18}$O($J=2-1$) line emission
toward the molecular-gas disk in M33 at an angular resolution of 7\farcs31 $\times$ 6\farcs50 (30\,pc $\times$ 26\,pc).
We combined the ACA 7\,m-array $^{12}$CO($J=2-1$) data with the IRAM 30\,m data to compensate for diffuse molecular-gas components.
The summary of this work is as follows:

\begin{enumerate}
\item
The ACA+IRAM combined $^{12}$CO($J=2-1$) map clearly depicts the cloud-scale molecular-gas structure over the M33 disk.
In addition, we detected a lot of $^{13}$CO($J=2-1$) sources which correspond to moderately dense molecular gas.

\item
We decomposed individual cloud components from the ACA+IRAM $^{12}$CO($J=2-1$) cube data employing \texttt{PYCPROPS}, and cataloged 848 molecular clouds with a mass range from $10^3$\,$M_{\odot}$ to $10^6$\,$M_{\odot}$.
We found that high-mass clouds ($M_{\rm CO} \geq 10^5\,M_{\odot}$) tend to associate with the $8\,\mu$m-bright sources in the spiral arm region,
while low-mass clouds ($M_{\rm CO} < 10^5\,M_{\odot}$) tend to be apart from such $8\,\mu$m-bright sources and to exist in the inter-arm region.

\item
We found that most of the molecular clouds in M33 show smaller velocity dispersions than the Galactic $R-\sigma_v$ relation at a given radius.
This is presumably due to low-surface density molecular clouds, which may be maintained even by the small turbulence.

\item
We found that a small IRAM GMC is identified as a single molecular cloud even in ACA+IRAM CO data, while a large IRAM GMC (typically its $M_{\rm CO}$ is larger than $3 \times 10^5 M_{\odot}$) can be resolved into multiple ACA+IRAM clouds.
The velocity dispersion of a large IRAM GMC is mainly dominated by the line-of-sight velocity difference between small clouds inside the GMC rather than the internal cloud velocity broadening.

\item
Based on the comparison between $M_{\rm CO}$ and $M_{\rm Vir}$ for ACA+IRAM clouds, we found that high-mass clouds in M33 are almost virialized.
This indicates that the high-mass star formation likely onsets within such high-mass clouds by the gravitational instability.

\end{enumerate}

\begin{acknowledgments}
We thank the anonymous referee for helpful comments, which significantly improved the manuscript.
This paper makes use of the following ALMA data: \dataset[ADS/JAO.ALMA\#2017.1.00461.S], \dataset[ADS/JAO.ALMA\#2018.A.00058.S], \dataset[ADS/JAO.ALMA\#2017.1.00901.S], and  \dataset[ADS/JAO.ALMA\#2019.1.01182.S].
ALMA is a partnership of ESO (representing its member states), NSF (USA) and NINS (Japan), together with NRC (Canada), NSC and ASIAA (Taiwan), and KASI (Republic of Korea), in cooperation with the Republic of Chile.
The Joint ALMA Observatory is operated by ESO, AUI/NRAO, and NAOJ.
This work is based on observations made with the {\it Spitzer Space Telescope}, which is operated by the Jet Propulsion Laboratory, California Institute of Technology, under a contract with NASA.
Data analysis was in part carried out on the Multi-wavelength Data Analysis System operated by the Astronomy Data Center (ADC), National Astronomical Observatory of Japan.
This work was supported by NAOJ ALMA Scientific Research grant Nos. 2022-22B and JSPS KAKENHI (grant Nos. JP18H05440, JP21H00049, and JP21K13962).
\software{CASA (v5.4.0; \citealt{mcmullin2007}), Astropy \citep{astropy2018}, APLpy (v1.1.1; \citealt{robitaille2012})}
\end{acknowledgments}

\appendix \label{app}

\section{Diffuse Emission Subtraction for IRAC 8\,$\mu$\lowercase{m} data}

In Section~\ref{sec:sf}, we measured 8\,$\mu$m flux as a proxy for star formation activity in M33.
Generally, star formation rates (SFRs) are estimated from H$\alpha$ (and also far-infrared emission such as 24$\mu$m) luminosities by assuming that all the H$\alpha$ emitting gas is ionized by the local star-forming region.
Although the typical size of an H\,{\sc ii} region is $\sim$ 0.1\,--\,10\,pc in the MW \citep[e.g.,][]{kennicutt1984, garay1999}, H$\alpha$ maps for nearby galaxies often show 100\,pc scale (or more) ionizing gas distributions.
The theoretical studies showed that clumpy density structures of ISM allow for larger escape fractions of ionizing radiation \citep[e.g.,][and references therein]{haffner2009}.
This indicates that the H$\alpha$ emitting gas is not necessarily ionized by the local star-forming region, and thus the diffuse components of H$\alpha$ emission should be considered for the estimation of SFRs.
Such diffuse components are also observed in the IRAC 8\,$\mu$m map that we used.

\begin{figure*}[ht!]
\epsscale{0.5}
\plotone{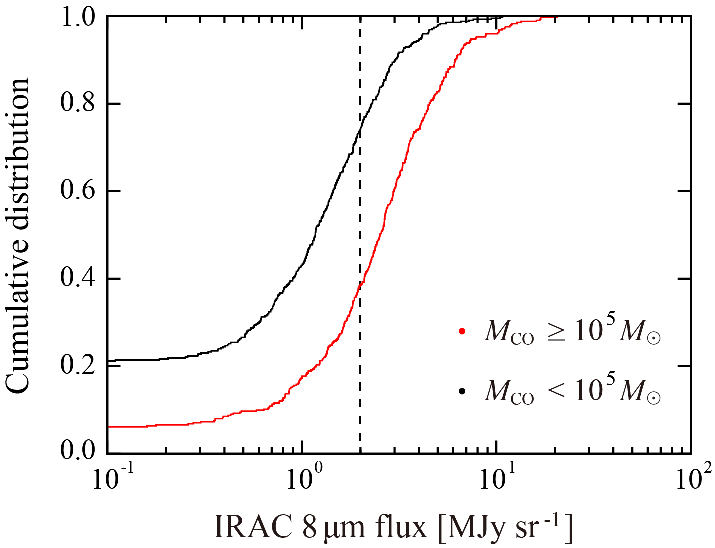}
\caption{Same as Figure~\ref{fig:cdf_8um}, but using the 8\,$\mu$m flux without diffuse components.
}\label{fig:cdf_8um_sub}
\end{figure*}

To extract the compact 8\,$\mu$m emission which directly reflects the star formation from the diffuse components, we applied $HIIphot$, an IDL software developed by \cite{thilker2000}.
Following the procedures in \cite{liu2011}, we subtracted the diffuse components from the 8\,$\mu$m map.
Figure~\ref{fig:cdf_8um_sub} shows the same as Figure~\ref{fig:cdf_8um}, but using the 8\,$\mu$m flux without diffuse components.
The general trend found in Figure~\ref{fig:cdf_8um} does not change; the $8\,\mu$m-bright sources are closely associated with high-mass clouds rather than low-mass clouds.

\end{document}